\documentclass[a4paper,fleqn,usenatbib]{mnras}

\usepackage{mathptmx}
\usepackage[T1]{fontenc}
\usepackage{ae,aecompl}

\usepackage[usenames,dvipsnames,svgnames,table]{xcolor}
\usepackage{graphicx}   % Including figure files
\usepackage{amsmath}    % Advanced maths commands
\usepackage{amssymb}    % Extra maths symbols
\newcommand{\kms}{\,$\rm km\;s^{-1}$}
\newcommand{\magasec}{\,mag\,arcsec$^{-2}$}
\newcommand{\msunpcsq}{\,M$_{\odot}$\,pc$^{-2}$} 
\newcommand{\galex}{\textit{GALEX}}
\newcommand{\mass}{\textsc{2Mass}}

\newcommand{\hopcat}{\textsc{Hopcat}}
\newcommand{\hipass}{\textsc{Hipass}}

\newcommand{\alfalfa}{\textsc{Alfalfa}}
\newcommand{\highmass}{H\textsc{i}ghMass}

\newcommand{\halogas}{\textsc{Halogas}}
\newcommand{\hix}{\textsc{Hix}}
\newcommand{\ds}{\textsc{Dark Sage}}
\newcommand{\control}{control}
\newcommand{\tirific}{TiRiFiC}
\newcommand{\miriad}{\textsc{miriad}}

\newcommand{\hi}{H\,\textsc{i}}
\newcommand{\htwo}{H$_2$}
\newcommand{\kband}{\textit{K}-band}

\newcommand{\rband}{\textit{R}-band}
\newcommand{\bjband}{$B_j$-band}

\newcommand\mycom[2]{\genfrac{}{}{0pt}{}{#1}{#2}}
\newcommand{\change}[1]{{#1}} % my changes, SeaGreen
\newcommand{\update}{{[Figure or Table updated]}}

\title[The HIX galaxy survey II]{The HIX galaxy survey II: HI kinematics of HI 
eXtreme galaxies}

\author[Lutz et al.]{
K. A. Lutz,$^{1, 2}$\thanks{E-mail: research@katha-lutz.de}  
V. A. Kilborn,$^{1}$ 
B. S. Koribalski,$^{2}$ 
B. Catinella,$^{3}$ \newauthor
G. I. G. J\'ozsa,$^{4, 5, 6}$  
O. I. Wong, $^{3}$
A. R. H. Stevens,$^{1, 3}$ 
D. Obreschkow,$^{3}$ \newauthor
H. D\'enes, $^{2, 7}$
\\
$^{1}$ Centre for Astrophysics and Supercomputing, Swinburne University of 
Technology, P.O. Box 218, Hawthorn, VIC 3122, Australia \\
$^{2}$ Australia Telescope National Facility, CSIRO Astronomy and Space 
Science, P.O. Box 76, Epping, NSW 1710, Australia\\
$^{3}$ International Centre for Radio Astronomy Research (ICRAR), M468, The 
University of Western Australia, \\
35 Stirling Highway, Crawley, WA 6009, Australia \\
$^{4}$ SKA South Africa Radio Astronomy Research Group, 3rd Floor, The Park, 
Park Road, Pinelands 7405, South Africa \\ 
$^{5}$ Rhodes Centre for Radio Astronomy Techniques \& Technologies, 
Department of Physics and Electronics, Rhodes University, \\
PO Box 94, Grahamstown 6140, South Africa \\
$^{6}$ Argelander-Institut f\"ur Astronomie, Auf dem H\"ugel 71, D-53121 Bonn, 
Germany \\
$^{7}$ Research School of Astronomy and Astrophysics, The Australian National 
University, Canberra, ACT 2611, Australia \\
}

\date{Accepted XXX. Received YYY; in original form ZZZ}

\pubyear{2018}

\begin{document}
\label{firstpage}
\pagerange{\pageref{firstpage}--\pageref{lastpage}}
\maketitle

\begin{abstract}
By analysing a sample of galaxies selected from the HI Parkes All Sky Survey 
(HIPASS) to contain more than 2.5 times their expected HI content based on 
their optical properties, we investigate what drives these HI eXtreme (HIX) 
galaxies to be so HI-rich. We model the \hi\ kinematics with 
the \textit{Ti}lted \textit{Ri}ng \textit{Fi}tting \textit{C}ode \tirific\ and 
compare the observed \hix\ galaxies 
to a control sample of galaxies from HIPASS as well as simulated galaxies 
built with the semi-analytic model \ds. We find that (1) 
\hi\ discs in \hix\ galaxies are more likely to be warped and more likely to 
host \hi\ arms and tails than in the \control\ galaxies, (2) 
the average \hi\ and average stellar column density of \hix\ galaxies is 
comparable to the \control\ sample, (3) \hix\ galaxies have higher \hi\ and 
baryonic specific angular momenta than \control\ galaxies, (4) most \hix\ 
galaxies live in higher-spin haloes than most \control\ galaxies. These results 
suggest that \hix\ galaxies are \hi-rich because they can support 
more \hi\ against gravitational instability due to their high specific angular 
momentum. The majority of the \hix\ galaxies inherits their high specific 
angular momentum from their halo. The \hi\ content of \hix\ galaxies might be 
further increased by gas-rich minor mergers.

This paper is based on data obtained with the Australia Telescope Compact 
Array (ATCA) through the large program C\,2705. 
\end{abstract}


\section{Introduction}
\label{sec:intro}
The gaseous and stellar content of galaxies is tightly related through the 
galactic gas cycle. Atomic hydrogen (\hi) condenses to form molecular gas 
(\htwo) clouds. These clouds are the birth places of stars. When comparing the 
amount of available \hi\ to the current star formation rate in local galaxies, 
\citet{Kennicutt1998} and \citet{Schiminovich2010} find that their \hi\ 
reservoirs would be consumed within $\approx 2$\,Gyr. Hence, galaxies need to 
replenish their gas reservoir in order to remain active starformers in the 
future (\citealp{Sancisi2008}, \citealp{SanchezAlmeida2014} and references 
therein). 

Gas--rich mergers and smooth accretion from the circumgalactic medium are 
suggested as avenues for gas replenishment \citep{White1978}. 
Observations of local galaxies do not find evidence for enough gas rich mergers 
to sustain star formation \citep{DiTeodoro2014,Sancisi2008,SanchezAlmeida2014}. 
This leads to the the conclusion that smooth accretion is the dominant channel 
of gas accretion. This might be the reaccretion of gas previously ejected by 
feedback mechanisms together with pristine halo gas, which is dragged along 
(\`a la the ``Galactic Fountain'', see e.\,g. 
\citealp{Oosterloo2007,Fraternali2011}). Cosmological 
simulations suggest accretion occurs through the cooling of hot halo gas or 
through the delivery of cold gas through filaments 
\citep{Birnboim2003,Keres2005,Dekel2006,vandeVoort2011a}. 
In the local Universe, gas-phase metallicity gradients / inhomogeneities 
\citep{Moran2012}, warps \citep{Roskar2010} and lopsided discs 
\citep{Bournaud2005} may be interpreted as observations of cosmological 
accretion but may also result from tidal interactions with other galaxies.

The \halogas\ survey \citep{Heald2011} has previously searched 
for signs of accretion in deep \hi\ observations of nearby galaxies (distance 
$<$ 11\,Mpc). Through detailed modelling of the \hi\ kinematics, the \halogas\ 
team has found a few high velocity clouds, thick \hi\ discs, and warps in their 
samples galaxies 
\citep{Gentile2013,Zschaechner2011,Zschaechner2012,deBlok2014}. In the 
Milky Way, high velocity clouds are thought to contribute to gas accretion 
\citep{Putman2012}. The thick disc component, which is usually lagging in 
rotation velocity with respect to the thin disc, is interpreted as a sign of 
the 
Galactic Fountain \citep{Oosterloo2007,Fraternali2011}. However, the total 
rate of detected \hi\ accretion in the \halogas\ observations is not sufficient 
to fuel star formation in their sample galaxies \citep{Heald2015}. 

The \hi\ eXtreme (\hix) galaxy survey examines a sample of \hi-rich 
galaxies to understand how they accumulate and maintain their gas reservoirs. 
In \citet{Lutz2017}, we found that \hix\ galaxies are less 
efficient at forming stars than a \control\ sample. The 
most extreme galaxy in the \hix\ sample (ESO075-G006) has built its massive 
\hi\ disc through a combination of a lower star formation 
efficiency ($\rm SFE_{HI} = SFR / M_{HI}$), due to a high specific baryonic 
angular momentum, and likely some accretion of pristine gas (as probed by 
gas-phase metallicity gradients). 

So the gas-rich galaxies of the \hix\ survey are not necessarily gas-rich due 
to recent gas accretion but could also be inefficient at using their available 
gas for star formation. Simple models describing the \hi\ based star formation 
efficiency  ($\rm SFE_{HI} = SFR / M_{HI}$) find a strong dependence of the SFE 
on the stability of the disc \citep{Wong2016}. \citet{Maddox2015} suggests that 
the upper envelope of the stellar -- \hi\ mass relation at high stellar masses 
is defined by the halo spin parameter. That is galaxies with a high \hi\ mass 
for their stellar mass tend to live in higher spin haloes. A high angular 
momentum can reduce the star formation efficiency in two ways: (1) accreted gas 
can not be transported to the denser, inner parts of the galaxy 
\citep{Kim2013,Forbes2014}, where the star formation efficiency would be higher 
\citep{Leroy2008}; (2) the disc is stabilised against star formation 
\citep{Toomre1964,Obreschkow2016}.

In this paper, we extend the analysis of the relation between the \hi\ content 
and kinematic properties to the entire \hix\ sample and an accompanying 
\control\ sample. We make use of observations of our sample galaxies with the 
Australia Telescope Compact Array (ATCA), which provide spatially resolved 
\hi\ distributions and kinematics. 

This article is structured as follow. In Sec. \ref{sec:data}, we 
discuss the selection of the \hix\ and \control\ samples and present the data 
used in this paper. In Sec. \ref{sec:results}, we present the results of the  
analysis of \hi\ kinematics and distribution. We then compare our results to 
the semi-analytic model \ds\ of galaxy evolution in Sec. \ref{sec:darksage}. 
The results are discussed in Sec. \ref{sec:discussion}. We then conclude in 
Sec. \ref{sec:conclude}. 

Throughout the paper we will assume a flat $\Lambda$CDM cosmology with the 
following cosmological parameters: H$_0$ = 70.0\,km\,Mpc$^{-1}$\,s$^{-1}$, 
$\Omega_{m}$ = 0.3. All velocities are used in the optical convention (cz).

\section{Samples and data}
\label{sec:data}
\subsection{Galaxy samples}
In this paper, we perform a kinematic analysis of galaxies in the \hix\ survey, 
which was first presented in \citet{Lutz2017}. 
\hix\ galaxies were selected as a subsample of the compilation presented by 
\citet{Denes2014}, who used their sample to calibrate scaling relations between 
\hi\ mass and optical luminosity. This parent sample consists of 1796\,galaxies 
from the \hipass\ catalogues \citep{Meyer2004,Koribalski2004}, which have 
reliable optical counterparts in \hopcat\ \citep{Doyle2005}. 

In our 2017 paper, \hix\ and \control\ samples included 13\,galaxies each. 
\hix\ galaxies were selected to lie at least $1.4\,\sigma$  above the 
\citet{Denes2014} scaling relation between \hi\ mass and absolute \rband\ 
magnitude. The \control\ sample has been selected from the same parent sample 
to lie within $\pm 0.7\,\sigma$ of the scaling relation. We exclude dwarf 
galaxies by restricting our sample to stellar masses greater 
than $\rm log\, M_{\star}\,[M_{\odot}] > 9.7$. In this paper, \control\ 
galaxies NGC\,4672, IC\,4366 and ESO462-G016 are excluded because the signal to 
noise ratio of their \hi\ data was too low.
% did not pass the quality check after 
% calibration (see Sec. \ref{sec:reduction}). 

For ESO208-G026 and IC\,4857 the last paper was inconclusive on 
whether these two galaxies are true \hix\ galaxies\change{, because in the 
\hi\ mass fraction vs. stellar mass plane, both galaxies were located within 
the 1\,$\sigma$ scatter of the \hipass\ parent sample (see below in 
Fig.~\ref{fig:mhi_vs_mstar})}. Upon examination of 
the \hopcat\ aperture that was used to measure the optical 
photometry of IC\,4857, it became apparent that a spiral arm to the east of the 
galaxy had not been included in the photometrical analysis. This led to an 
underestimation of the \rband\ magnitude and subsequently an overestimation of 
the \hi-richness. IC\,4857 is therefore reclassified as a \control\ 
galaxy. As will be seen in the analysis of this paper, the properties 
of the \hi\ disc of IC\,4857 are furthermore similar to the remaining \control\ 
sample and not to the \hix\ sample. \change{Neither the analysis of the 
optical/near-infrared photometry nor of the resolved \hi\ observations of  
ESO208-G026 were able to explain why this galaxy is a clear outlier on
the \citet{Denes2014} relation but not as much on the \hi\ mass 
fraction vs. stellar mass plane. Therefore,} ESO208-G026 remains in the \hix\ 
sample. 

\subsection{Ancillary data}

Information on the stellar content of the \hix\ and \control\ galaxies is based 
on data from the 2\,Micron All Sky Survey (\mass) and the associated Extended 
Source Catalogue (2MASX, \citealp{Skrutskie2006}). Stellar masses are 
calculated from the total 2MASX \kband\ magnitude following equ. 3 of 
\citet{Wen2013}:
\begin{equation}
    \log M_{\star} [{\rm M_{\odot}}] = -0.498 + 1.105 \times 
\log L_{K} [{\rm L_{\odot}}].
\label{equ:mstar}
\end{equation}
Only 1478 out of the 1796\,galaxies in the parent sample have a cross match in 
the 2MASX catalogue (82\,per\,cent). These ``missing'' galaxies are excluded 
from the analysis.

In section \ref{sec:stabel}, we will consider the atomic gas mass $M_{A}$ 
and the baryonic mass $M_{B}$. We define these masses as follows:
\begin{equation}
    M_{A} = M_{HI + He} = 1.35 M_{HI}
\end{equation}
i.\,e. the \hi\ mass with a correction factor to account for Helium (e.\,g. 
\citealp{Obreschkow2016}) and
\begin{equation}
\label{equ:mbary}
    M_{B} = M_{A} + M_{\star}.
\end{equation}
where  M$_{\star}$ is the stellar mass. We neglect the molecular gas mass in 
this calculation of the baryonic mass, because the ISM of \hix\ and \control\ 
galaxies is dominated by atomic gas and estimated molecular gas masses (e.\,g. 
using scaling relations by \citet{Catinella2010} and \citet{Saintonge2011}) are 
an order of magnitude smaller than atomic gas masses. 

\subsection{Resolved HI data from ATCA}
\label{sec:reduction}
For this paper we focus on the \hi\ interferometric data of the \hix\ and 
\control\ sample. All data has been obtained with the Australia Telescope 
Compact Array (ATCA). While the majority of the observations for the \hix\ 
galaxies have been carried out as part of the large program C\,2705, the data 
of the \control\ sample  has been collected from the Australia Telescope 
Online Archive (ATOA)\footnote{http://atoa.atnf.csiro.au/}. Table \ref{tab:obs} 
summarises all ATCA observations that will be used in this paper. 

\begin{table*}
\begin{tabular}[H]{l || c c c c c c }
     ID       & 1.5km Array & 750m Array & EW Array & $\rm t_{O\,S}$ [h] & 
Phase Calibrator & $\rm f_{Cen}$ [MHz] \\
    (1) & (2) & (3) & (4) & (5) & (6) & (7)  \\  \hline \hline
    ESO111-G014& 2012-12-19         & 2013-01-20         & 2013-01-16         
        & 24.6  & 2355-534 & 1384 \\ 
    ESO243-G002& 2015-06-15         & \dots              & 2012-11-08   
        & 17.9  & 0022-423 & 1378 \\ 
    NGC\,289   & 2002-07-06$^{(a)}$ & 2015-09-07         & 2015-08-28        
        & 27.7  & 0022-423 / 0042-357 & 1413 \\ 
    ESO245-G010& 2014-11-16         & 2015-02-24         & 2014-11-29         
        & 20.9  & 0201-440 & 1393 \\ 
    ESO417-G018& 2014-11-14         & 2015-09-08         & 
$\mycom{2014-11-30}{2012-05-07}$      
        & 20.1  & 0237-233 / 0153-410 & 1398 \\ 
    ESO055-G013& 2014-11-15         & 2015-02-24         & 2014-12-01         
        & 21.6  & 0252-712 & 1385 \\ 
    ESO208-G026& 2014-11-17         & 2015-09-10         & 2015-08-29         
        & 29.5  & 0823-500 & 1406 \\ 
    ESO378-G003& 2015-06-13         & 2015-09-09         & 2015-08-31         
        & 28.3  & 1144-379 & 1406 \\ 
    ESO381-G005& 2015-06-14         & 2015-09-07         & 2015-08-31         
        & 28.4  & 1232-416 & 1394 \\ 
    ESO461-G010& 2014-11-14         & 2015-02-23         & \dots               
        & 13.6  & 1921-293 & 1389 \\ 
    ESO075-G006& 2012-12-20         & 2013-01-19         & 2012-11-08         
        & 20.5  & 1934-638 & 1370 \\ 
    ESO290-G035& 2014-11-15         & 2015-02-14         & $\mycom{2014-11-29} 
{2012-05-08}$
        & 20.3  & 2333-528 & 1392 \\ \hline
    
    NGC\,685   & 1995-10-29$^{(b)}$ & 1995-09-30$^{(b)}$ & \dots               
        & 20.0 & 0407-658 & 1414 \\ 
    ESO121-G026& 2014-05-13$^{(c)}$ & \dots               & \dots               
        & 10.9 & 0539-530 & 1396 \\ 
    ESO123-G023& 2001-08-04$^{(d)}$ & 2000-05-10$^{(d)}$ & \dots               
        & 20.0 & 0727-365 & 1407 \\ 
    NGC\,3001  & 2014-05-12$^{(c)}$ & \dots               & \dots               
        & 11.1 & 0919-260 & 1396 \\ 
    ESO263-G015& 2001-03-15$^{(e)}$ & \dots               & \dots               
        & 10.8 & 1039-47  & 1408 \\ 
    NGC\,3261  & 1997-05-01$^{(f)}$ & 1997-06-09$^{(f)}$ & 1997-04-02$^{(f)}$
        & 30.0 & 0823-500 / 1215-457 & 1407 \\ 
    NGC\,5161  & 2014-05-14$^{(c)}$ & \dots               & \dots               
        & 10.5 & 1255-316 & 1396 \\ 
    ESO383-G005& 2000-05-20$^{(g)}$ & \dots                  & \dots           
        & 9.8  & 1320-446 & 1403 \\ 
    IC\,4857   & 2015-06-17         & 2015-09-05         & 2012-06-06         
        & 20.4  & 1934-638 & 1399 \\ 
    ESO287-G013& 2001-10-28$^{(d)}$ & 2000-05-09$^{(d)}$ & \dots               
        & 20.6 & 2106-413 & 1408 \\ 
    ESO240-G011& 2001-07-28 / 29$^{(d)}$ & 2000-05-11$^{(d)}$ & \dots           
    
        & 19.9 & 2326-477 & 1407 \\ 

\end{tabular}
    \caption{The ATCA \hi\ observations used in this paper. Observations listed
        above the horizontal line belong to \hix\ galaxies, below to \control\ 
        galaxies. \textit{Column} (1): Galaxy ID as defined in 
\citet{Dreyer1888,Dreyer1910,Lauberts1989}. \textit {Column} (2), (3) 
        and (4): Date of observation in any 1.5\,km, 750m and EW array 
        configuration. \textit{Column} (5): On source observation time in 
        hours. \textit{Column} (6): Phase Calibrator PKS\,[ID number] (based on 
the Parkes Catalogue \citep{Wright1990} and its predecessors). 
        \textit{Column} (7): Central frequency of band in MHz. Most 
        \hix\ galaxy observations have been observed in the project C\,2705. 
        More observations are taken from the following projects: $^{(a)}$ 
        C\,819, $^{(b)}$ C\,473, $^{(c)}$ C\,2921, $^{(d)}$ C\,885, $^{(e)}$ 
        C\,869, $^{(f)}$ C\,633 and $^{(g)}$ C\,801.}
\label{tab:obs} 
\end{table*}

Since 2009, the ATCA is equipped with the Compact Array Broadband Backend 
(CABB, \citealp{Wilson2011}). All observations after 2009 used CABB but 
depending on the observing projectr, CABB has been used in two 
different modes: CFB~1M-0.5k (0.5\,kHz resolution) and CFB~64M-34k mode 
(34\,kHz resolution, used in project C\,2921). All observations prior to 2009 
were obtained with the previous correlator (16\,kHz resolution). 

The methodology of the data reduction is similar for all three set-ups, and a 
semi-automated \miriad\ pipeline is used in all cases. Firstly, radio frequency 
interference (e.\,g. signal from GPS Satellites) is 
removed. A semi-automated pipeline conducts bandpass, flux, and phase 
calibration using the \miriad\ \citep{Sault1995} tasks \textsc{gpcal} and 
\textsc{mfboot}. For 
bandpass and flux calibration the standard calibration source 
PKS\,1934-638 has been observed before or after each galaxy observation. The 
phase calibrator is given in Tab. \ref{tab:obs} and has been visited regularly 
during the galaxy observations. The calibration is then applied to the galaxy 
data. In the final step of the pipeline, a 
first-order polynomial baseline is subtracted from each galaxy 
observation. We consider the results of this procedure satisfactory, once the 
phase of the phase calibrator is tightly scattered around 0 across the 
observing period and the amplitudes of the phase and bandpass calibrators are 
in agreement with the ATCA calibrator data 
base.\footnote{https://www.narrabri.atnf.csiro.au/calibrators/} 

A second pipeline combines all available observations of one galaxy from 
different 
array configurations in the Fourier transformation. For this step, the \miriad\ 
task \textsc{invert} is used. The obtained dirty data cube is then cleaned 
(\textsc{clean}), restored (\textsc{restor}) and primary beam corrected 
(\textsc{linmos}). Mosaicking is necessary in one case, where data has been 
compiled from different observing projects (NGC\,289). In this case cleaning is 
carried out with \textsc{mossdi} and no primary beam correction is necessary. 

Moment 0, 1 and 2 maps are created with the \miriad\ 
task \textsc{moment}. A three-sigma clipping is applied. Furthermore, values in 
the moment 1 maps are restricted such that only velocities that can be 
measured with our correlator set-up are included. Both options reduce the noise 
in the resulting maps. Moment 1 and 2 maps are then masked to regions where the 
column density in the moment 0 maps is larger than $\rm 0.4\ and\ 0.8 \times  
10^{20}\,cm^{-2}$ for \hix\ and \control\ sample, respectively. In addition,
moment 0 maps without the three-sigma clipping are created. 

\change{Within the errors, the \hi\ masses as measured from the ATCA \hi\ data 
cubes and the \hi\ masses as measured from the integrated flux from the 
\hipass\ catalogues agree for all \hix\ galaxies and for half of the 
\control\ galaxies. In the appendices, a \hipass\ spectrum is given, which has 
been re-measured from the \hipass\ data cubes. Due to different baselines, 
the \hi\ masses as measured from the ATCA \hi\ data cubes and as re-measured 
from \hipass\ data cubes only agree within the errors for 83 and 43\,per\,cent 
of the \hix\ and the \control\ galaxies, respectively. }

The different column density limits \change{and flux recovery percentages }are 
due to the fact that the \hix\ observations are, in general, more sensitive than 
the \control\ galaxy observations because of:
\begin{itemize}
 \item longer total on-source integration times for the \hix\ galaxies.
 \item higher system temperature of the old correlator, which was used for 
most of the \control\ sample observations.
\end{itemize}

For \change{most }results in this paper, the different limits do not pose a 
problem because:
\begin{enumerate}
    \item \change{all extended features in the \hix\ galaxies that are 
relevant in our analysis are also detected at column densities 
$\rm > 0.8 \times 10^{20}\,cm^{-2}$. }
    \item the \hi\ radius is measured at $\rm 1\,M_{\odot}\,pc^{-2} = 
1.2 \times 10^{20}\,cm^{-2}$ (see Sec. \ref{sec:radprof})\change{, well above 
both detection limits}.
\end{enumerate}
\change{It is, however, to be noted that the observations of \control\ 
galaxies generally include fewer short baselines than the observations of 
\hix\ galaxies. Hence, some diffuse features in \control\ galaxies may be missed 
in this analysis. This problem may only be resolved with significant amounts of 
new data. For example, arms and tails as found in \hix\ galaxies ESO381-G005 
and ESO208-G026 are only detected once we at least combine observations from the 
1.5\,km and the 750\,m array configuration. When only using 1.5\,km array 
configuration observations, we are able to detect the tail in ESO378-G003 but 
not the associated cloud in this galaxy or any arms in ESO381-G005 and 
ESO208-G026 (see Sec. \ref{sec:morph} for more details). The combination of 
observations of the 1.5\,km and the 750\,m array configuration is only 
available for six out of eleven \control\ galaxies. }

Table \ref{tab:cubes} summarises the characteristics of the final data cubes of 
each galaxy. The rms of the data cubes is around 1 
to 3\,mJy\,beam$^{-1}$. Beam sizes range between 65 by 43\,arcsec and 25 by 
18\,arcsec. 

\begin{table}
\begin{tabular}[H]{l || c c c }
ID & $\theta_{1}$ & $\theta_{2}$ & rms  \\
~ &  [\arcsec] &  [\arcsec] & [mJy\,beam$^{-1}$] \\
		(1)& (2) & (3) & (4)  \\  \hline \hline
    ESO111-G014         & 52.66 & 40.58 & 1.7 \\
    ESO243-G002         & 31.70 & 22.95 & 1.5 \\
     NGC\,289$^a$       & 55.98 & 25.64 & 1.4 \\
    ESO245-G010         & 51.48 & 31.30 & 1.7 \\
    ESO417-G018         & 66.09 & 33.33 & 1.6 \\
    ESO055-G013         & 43.08 & 33.54 & 2.0 \\
    ESO208-G026         & 59.24 & 43.49 & 1.3 \\
    ESO378-G003         & 64.23 & 41.09 & 1.3 \\
    ESO381-G005         & 65.76 & 43.50 & 1.5 \\
    ESO461-G010$^b$     & 59.62 & 29.38 & 2.7 \\
    ESO075-G006         & 60.52 & 35.22 & 2.0 \\
    ESO290-G035$^b$     & 58.32 & 28.98 & 2.1 \\   \hline
    NGC\,685            & 34.58 & 32.48 & 1.9 \\
    ESO121-G026$^c$     & 25.42 & 18.91 & 0.6 \\
    ESO123-G023         & 35.74 & 28.94 & 2.3 \\
    NGC\,3001$^c$       & 38.49 & 20.97 & 1.3 \\
    ESO263-G015         & 25.53 & 22.60 & 3.0 \\
    NGC\,3261           & 56.50 & 39.31 & 2.7 \\
    NGC\,5161$^c$       & 36.59 & 20.41 & 1.3 \\
    ESO383-G005         & 40.83 & 22.02 & 2.8 \\
       IC\,4857         & 40.16 & 26.92 & 1.1 \\
    ESO287-G013         & 38.00 & 30.33 & 2.0 \\
    ESO240-G011         & 29.32 & 22.87 & 3.0 \\
\end{tabular}

    \caption{Characteristics of the final data cubes for \hix\ (above 
horizontal line) and \control\ 
galaxies (below horizontal line). \textit{Column} (1): Galaxy ID. \textit 
{Column} (2): Beam major axis in arcsec. \textit{Column} (3):  Beam 
minor axis in arcsec. \textit{Column} (4): RMS in the final data cube. The 
weighting is set by the robust parameter in \miriad's \textsc{invert} task. A 
weighting of 2.0 indicates natural weighting and a weighting of -2.0 
corresponds to uniform weighting. For most cubes a weighting with robust=0.5 
was chosen except for $^a$ and $^b$ where robust = 2.0 and 0.0 was chose 
respectively. $^c$ indicates that the data cube has a velocity 
channel width of 6\kms instead of 4\kms.}
\label{tab:cubes}
\end{table}
% \newpage
\subsection{TiRiFiC modelling}
\label{sec:tiri}

\begin{table*}
    \begin{tabular}[H]{l || c c c c c c }
        ID & $\rm V_{rot}$ & $R_{25}$ & PA$_0$ & INCL$_0$ & $\Delta$ PA & 
        $\Delta$ INCL  \\
        ~ &  [$\rm km~s^{-1}$] & [kpc] &  [deg] & [deg] &  [deg] & [deg] \\
        (1)& (2)   & (3) & (4) & (5) & (6) & (7)   \\  \hline \hline
        ESO111-G014 & 204.4 & 23.5 & 337.9 & 55.6 & 12.4 & 15.9 \\
        ESO243-G002 & 166.5 & 16.4 & 349.9 & 47.7 & 0.0 & 0.0 \\
        NGC0289     & 191.0 & 11.0 & 135.6 & 48.2 & 18.6 & 24.1 \\
        ESO245-G010 & 192.4 & 23.5 & 25.5 & 72.8 & 7.5 & 8.3 \\
        ESO417-G018 & 173.9 & 19.9 & 357.0 & 57.8 & 3.0 & 5.1 \\
        ESO055-G013 & 196.8 & 8.1  & 51.1 & 26.4 & 48.2 & 3.2 \\
        ESO208-G026 & 143.1 & 6.6  & 286.0 & 63.1 & 16.9 & 34.1 \\
        ESO378-G003 & 134.8 & 10.3 & 354.8 & 62.7 & 15.2 & 4.7 \\
        ESO381-G005 & 122.4 & 11.7 & 313.4 & 55.3 & 85.3 & 8.5 \\
        ESO461-G010 & 153.9 & 17.3 & 77.4 & 83.9 & 21.8 & 5.5 \\
        ESO075-G006 & 217.5 & 21.0 & 315.9 & 39.8 & 35.6 & 5.3 \\
        ESO290-G035 & 181.6 & 23.7 & 338.2 & 81.5 & 21.3 & 1.8 \\  \hline
        NGC\,685    & 127.7 & 9.1  & 100.3 & 36.0 & 0.0 & 0.0 \\
        ESO121-G026 & 200.7 & 11.8 & 294.4 & 44.9 & 0.0 & 0.0 \\
        ESO123-G023 & 157.8 & 12.8 & 288.8 & 90.0 & 22.0 & 31.1 \\
        NGC\,3001   & 249.6 & 14.3 & 185.4 & 51.1 & 16.0 & 3.0 \\
        ESO263-G015 & 159.3 & 14.7 & 289.2 & 90.0 & 3.0 & 9.7 \\
        NGC\,3261   & 349.4 & 17.3 & 256.7 & 42.3 & 37.9 & 9.9 \\
        NGC\,5161   & 175.0 & 19.6 & 250.7 & 63.3 & 13.0 & 15.9 \\
        ESO383-G005 & 207.0 & 22.7 & 310.3 & 85.3 & 11.7 & 3.4 \\
        IC\,4857    & 154.6 & 17.1 & 33.8 & 53.5 & 28.4 & 19.0 \\
        ESO287-G013 & 175.5 & 19.0 & 241.4 & 81.0 & 0.0 & 0.0 \\
        ESO240-G011 & 266.0 & 31.6 & 127.1 & 87.0 & 0.0 & 0.0 \\
        
    \end{tabular}

    \caption{Results of the \tirific\ modelling. The final models of galaxies 
        marked with $^{*}$ are FLAT models, all other galaxies are modelled 
        with WARP 
        models. \textit{Column} (1): The galaxy ID. \textit{Column} (2): The 
        rotation 
        velocity in $km~s^{-1}$. \change{\textit{Column} (3): the \bjband\ 
            25\magasec\ isophotal radius in kpc.  \textit{Columns} (4) and (5): 
            the 
            position angle and inclination values fitted to the rings inside of 
            R$_{25}$. 
            \textit{Columns} (6) and (7): the difference between the 
            largest and the smallest position angle and inclination values 
            respectively. }}
    \label{tab:tiri}
\end{table*}

We use the TIlted RIng FIitting Code \tirific\ \citep{Jozsa2007} to analyse 
the kinematic properties of \hix\ and \control\ galaxies. The 
tilted ring fitting method has first been introduced by \citet{Rogstad1974} and 
allows one to measure rotation velocity, position angle, inclination and more 
properties in concentric annuli.  

In the case of the \hix\ and \control\ sample we consider the following two 
models:
\begin{itemize}
  \item a flat disc (FLAT). In this case only the surface brightness and the 
rotation curve are allowed to vary with radius. All other parameters like 
inclination, position angle, centre of rotation, velocity dispersion, thickness 
of the disc and systemic velocity are the same in all rings.
  \item a warped disc with radius-dependent inclination and position angle 
(WARP). 
\end{itemize}
The inclination and position angle \change{within the optical 
\bjband\ 25\magasec\ 
isophotal radius \citep{Lauberts1989} are }kept constant regardless of the 
model to be fitted, 
because the optical discs of both the \hix\ and the \control\ sample do not 
appear warped upon visual inspection (see e.\,g. the approach to model 
NGC\,4414 by \citealp{deBlok2014}). \change{This means that position angle and 
inclination are fitted simultaneously for all rings within the optical 
25\magasec\ isophotal radius.}

% \newpage

The width of the rings is of the order of the beam size, and rings with smaller
widths are introduced at small radii. The number of rings is chosen such that 
the outermost ring is the first one without emission. 

To avoid over-interpreting our data, we did not fit more parameters like, 
for example, a vertical lag in 
rotation velocity, which might indicate an active Galactic Fountain (as 
described in Sec. \ref{sec:intro}). Our sample galaxies are located 
at distances between 18\,Mpc and 148\,Mpc, which is two to ten times further 
away than the \halogas\ galaxies. Therefore, one resolution element covers a 
larger fraction of the galaxy than was the case for \halogas\ galaxies.  The 
\tirific\ modelling does, however, include the geometrical ``Z0'' parameter, 
which is the thickness of the disc. 

For every galaxy in the \control\ and \hix\ sample, we chose a final \tirific\ 
model by:
\begin{itemize}
 \item visual inspection of channel maps of the input data cube and the output 
model cube.
 \item aiming for radially smooth variations (if any) of rotation velocity, 
inclination and position angle.
  \item minimising the free parameters that are fitted. That means, if a 
flat disc model and a warped disc model produce similarly small residuals we 
will chose the flat disc model. 
\end{itemize}

\newpage
As output, \tirific\ provides a model data cube and a table with the radial 
profiles of the fitted parameters.

\subsection{Radial profiles}
\label{sec:radprof}

\begin{table*}
    \begin{tabular}[H]{l || c c c c c c c c c c}
        ID & RA & D & $\rm \log\,M_{\star} $ & $\rm \log\,M_{HI, HIP} 
        $ & $\rm \log\,M_{HI, ATCA} $ & $\rm R_{HI}$ & $\rm 
        R_{eff}$ & $\rm j_{B}$ & $\lambda$ & $\rm \log\,q$ \\  
        ~ & DEC & Mpc & $\rm [M_{\odot}]$ & $\rm [M_{\odot}]$ & $\rm 
        [M_{\odot}]$ & 
        $\rm[kpc]$ & $\rm  [kpc]$ & $\rm [kpc~km~s^{-1}]$ & ~ & ~ \\  
        
        (1)        & (2)        & (3)    & (4)  & (5) & (6) & (7) & (8) & (9)   
        & (10) & (11) \\ \hline
        ESO111-G014& 00:08:18.8 & 111.98 &  10.5& 10.7& 10.6& 49.6& 6.0 & 
        4125.6 & 0.06 & -0.84  \\
        ~          & -59:30:56  & ~      & ~    & ~   & ~   & ~   & ~   & 
        ~      & ~    &~       \\
        ESO243-G002& 00:49:34.5 & 128.87 &  10.7& 10.8& 10.7& 55.6& 7.2 & 
        3523.8 & 0.05 & -0.82  \\
        ~          & -46:52:28  & ~      & ~    & ~   & ~   & ~   & ~   & 
        ~      & ~    &~       \\
        NGC\,289   & 00:52:42.4 & 23.06  &  10.5& 10.4& 10.5& 86.9& 2.9 & 
        5453.1 & 0.10 & -0.59  \\
        ~          & -31:12:21  & ~      & ~    & ~   & ~   & ~   & ~   & 
        ~      & ~    &~       \\
        ESO245-G010& 01:56:44.5 & 81.69  &  10.5& 10.3& 10.5& 50.9& 6.2 & 
        4092.4 & 0.07 & -0.63  \\
        ~          & -43:58:21  & ~      & ~    & ~   & ~   & ~   & ~   & 
        ~      & ~    &~       \\
        ESO417-G018& 03:07:13.2 & 67.35  &  10.3& 10.6& 10.4& 45.5& 5.7 & 
        3522.7 & 0.08 & -0.74  \\
        ~          & -31:24:03  & ~      & ~    & ~   & ~   & ~   & ~   & 
        ~      & ~    &~       \\
        ESO055-G013& 04:11:43.0 & 104.98 &  10.2& 10.5& 10.3& 41.0& 2.3 & 
        3749.8 & 0.10 & -0.68  \\
        ~          & -70:13:59  & ~      & ~    & ~   & ~   & ~   & ~   & 
        ~      & ~    &~       \\
        ESO208-G026& 07:35:21.1 & 39.72  &  9.8 & 9.8 & 9.9 & 28.8& 1.8 & 
        1993.0 & 0.10 & -0.45  \\
        ~          & -50:02:35  & ~      & ~    & ~   & ~   & ~   & ~   & 
        ~      & ~    &~       \\
        ESO378-G003& 11:28:04.0 & 40.56  &  10.1& 10.3& 10.3& 44.3& 4.8 & 
        3196.7 & 0.09 & -0.51  \\
        ~          & -36:32:34  & ~      & ~    & ~   & ~   & ~   & ~   & 
        ~      & ~    &~       \\
        ESO381-G005& 12:40:32.7 & 80.04  &  10.1& 10.4& 10.4& 37.8& 3.3 & 
        3388.2 & 0.08 & -0.47  \\
        ~          & -36:58:05  & ~      & ~    & ~   & ~   & ~   & ~   & 
        ~      & ~    &~       \\
        ESO461-G010& 19:54:04.4 & 98.26  &  10.1& 10.7& 10.3& 23.7& 4.0 & 
        6377.8 & 0.17 & -0.31  \\
        ~          & -30:29:03  & ~      & ~    & ~   & ~   & ~   & ~   & 
        ~      & ~    &~       \\
        ESO075-G006& 21:23:29.5 & 153.94 &  10.6& 10.8& 10.9& 94.0& 3.8 & 
        9764.0 & 0.11 & -0.76  \\
        ~          & -69:41:05  & ~      & ~    & ~   & ~   & ~   & ~   & 
        ~      & ~    &~       \\
        ESO290-G035& 23:01:32.5 & 84.50  &  10.5& 10.4& 10.5& 36.3& 5.3 & 
        4117.7 & 0.08 & -0.74  \\ 
        ~          & -46:38:47  & ~      & ~    & ~   & ~   & ~   & ~   & 
        ~      & ~    &~       \\ \hline
        NGC\,685   & 01:47:42.8 & 17.97  & 9.8  & 9.6 &  9.6& 16.1& 5.4 & 821.0 
        & 0.05 & -0.70   \\
        ~          & -52:45:42  & ~      & ~    & ~   & ~   & ~   & ~   & 
        ~      & ~    &~       \\ 
        ESO121-G026& 06:21:38.8 & 29.64  & 10.3 & 10.0& 9.6 & 21.5& 3.4 & 
        1041.3 & 0.04 & -1.14  \\
        ~          & -59:44:24  & ~      & ~    & ~   & ~   & ~   & ~   & 
        ~      & ~    &~       \\ 
        ESO123-G023& 07:44:38.9 & 38.63  & 10.0 & 9.9 & 10.0& 30.0& 3.7 & 
        2193.1 & 0.09 & -0.59  \\
        ~          & -58:09:13  & ~      & ~    & ~   & ~   & ~   & ~   & 
        ~      & ~    &~       \\ 
        NGC\,3001  & 09:46:18.7 & 32.14  & 10.7 & 10.0&  9.8& 30.9& 3.9 & 
        1469.1 & 0.03 & -1.21  \\
        ~          & -30:26:15  & ~      & ~    & ~   & ~   & ~   & ~   & 
        ~      & ~    &~       \\ 
        ESO263-G015& 10:12:19.9 & 33.13  & 10.3 & 9.6 & 9.8 & 22.5& 5.9 & 
        2447.9 & 0.08 & -0.72  \\
        ~          & -47:17:42  & ~      & ~    & ~   & ~   & ~   & ~   & 
        ~      & ~    &~       \\ 
        NGC\,3261  & 10:29:01.5 & 33.59  & 10.8 & 10.2& 10.0& 30.8& 4.5 & 
        2252.0 & 0.04 & -1.14  \\
        ~          & -44:39:24  & ~      & ~    & ~   & ~   & ~   & ~   & 
        ~      & ~    &~       \\ 
        NGC\,5161  & 13:29:13.9 & 32.23  & 10.6 & 10.3& 10.3& 41.6& 6.0 & 
        2294.8 & 0.05 & -0.91  \\
        ~          & -33:10:26  & ~      & ~    & ~   & ~   & ~   & ~   & 
        ~      & ~    &~       \\ 
        ESO383-G005& 13:29:23.6 & 50.20  & 10.6 & 9.7 & 9.7 & 23.8& 4.1 & 
        2035.6 & 0.05 & -1.01  \\
        ~          & -34:16:17  & ~      & ~    & ~   & ~   & ~   & ~   & 
        ~      & ~    &~       \\ 
        IC\,4857   & 19:28:39.2 & 66.50  & 10.5 & 10.5& 10.0& 35.1& 7.9 & 
        2095.4 & 0.05 & -0.78  \\
        ~          & -58:46:04  & ~      & ~    & ~   & ~   & ~   & ~   & 
        ~      & ~    &~       \\ 
        ESO287-G013& 21:23:13.9 & 38.72  & 10.5 & 10.2& 10.2& 38.3& 4.4 & 
        4095.8 & 0.07 & -0.86  \\
        ~          & -45:46:23  & ~      & ~    & ~   & ~   & ~   & ~   & 
        ~      & ~    &~       \\ 
        ESO240-G011& 23:37:49.9 & 40.12  & 10.8 & 10.5& 9.8 & 29.2& 5.7 & 
        1907.1 & 0.05 & -0.92  \\
        ~          & -47:43:41  & ~      & ~    & ~   & ~   & ~   & ~   & 
        ~      & ~    &~       \\ 
        
    \end{tabular}
    \caption{Basic properties and measurements for the \hix\ (above the 
        horizontal line) and \control\ galaxy sample (below the horizontal 
        line). 
        \textit{Column} (1): Galaxy ID. \textit {Column} (2): 2MASX coordinates 
        of 
        the stellar centre. \textit {Column} (3): distance in Mpc. 
        \textit{Column} (4): 
        Stellar masses as calculated from the 2MASX \kband\ luminosity. 
        \textit{Column} (5): 
        \hi\ mass as measured from \hipass\ data 
        cubes. \textit{Column} (6): \hi\ mass as measured from ATCA data cubes 
        in this 
        work. \textit{Column} (7): The \hi\ radius as measured at the 
        1\,\msunpcsq\ 
        isophote in kpc. \textit{Column} (8): The 2MASX \kband\ half-light 
        radius in 
        kpc. \textit{Column} (9): The baryonic specific angular momentum in 
        units of 
        $\rm [kpc~km~s^{-1}]$. 
        \textit{Column} (10): The halo spin parameter inferred from the 
        baryonic 
        specific angular momentum. 
        \textit{Column} (11): The global stability parameter q. }
    \label{tab:props} 
\end{table*}

The radial profiles of the \hi\ column density, \hi\ mass and stellar mass are 
measured using the radial profiles of inclination and position angle from 
\tirific. In a first step we extrapolate inclination and position angle between 
the \tirific\ rings. Then concentric, elliptical annuli are defined. The width 
of these annuli is of the order of the pixel size. Inclination and position 
angle of each annulus are taken from the extrapolated \tirific\ profiles 
according to their radius. Within each elliptical annulus, we measure the total 
\hi\ mass, the total stellar mass and the average \hi\ column density. \hi\ 
mass 
and column density profiles are measured from the non-clipped moment 0 maps 
and stellar mass profiles from the \mass\ \kband\ 
images. The \mass\ luminosities in every annulus are converted to stellar 
masses again using equ. 3 of \citet{Wen2013} (see Equ. \ref{equ:mstar}).
Stellar masses are only determined in annuli, which are smaller than the 
aperture within which \mass\ measured the integrated \kband\ magnitude. 

\tirific\ also provides rotation velocities for each fitted ring. These 
rotation velocities are fitted with a rotation curve of the functional form:
\begin{equation}
\label{equ:vrot}
 V_{\rm rot} (r) = V_{\rm flat} \cdot \left[ 1 - 
\exp\left(\frac{-r}{l_{\rm flat}}\right) 
\right]
\end{equation}
(\citealp{Leroy2008} and references therein). $V_{\rm flat}$ is the circular 
velocity in the flat part of the rotation curve and $l_{\rm flat}$ describes 
the radius at which the rotation velocity flattens.

The \hi\ radius, $R_{\rm HI}$, is measured from the \hi\ column density 
profiles. $R_{\rm HI}$ is defined to be the radius at which the \hi\ column 
density drops to 1\msunpcsq. Before $R_{\rm HI}$ is measured, \hi\ column 
densities are deprojected by multiplying with the cosine of the inclination 
$\cos\,i$. Therefore, in more edge-on galaxies, the measurement of $R_{\rm HI}$ 
is more sensitive to inclination variations than in more face-on galaxies. The 
radius is determined by linearly extrapolating the two column density profile 
data points just above and just below 1\msunpcsq. Following 
\citet{Wang2014,Wang2016}, $R_{\rm HI}$ is then corrected for beam smearing 
effects. This reduces  $R_{\rm HI}$ on average by 1\,per\,cent and at most by 
4\,per\,cent

The combination of the stellar or \hi\ mass profiles with the rotation curve 
allows one to calculate the stellar or \hi\ specific angular momentum of the 
disc following
\begin{equation}
\label{equ:jb}
 j = \frac{\sum_{i} M_{i} \times V_{rot, i} \times r_{i}} {\sum_{i} M_{i}}.
\end{equation}
\citep{Obreschkow2014, Obreschkow2016}, where $M_{\rm i}$ is the mass of ring i 
with radius $r_{\rm i}$ rotating at velocity $V_{\rm rot, i}$. The baryonic 
specific angular momentum is the mass-weighted average of the stellar and \hi\ 
specific angular momenta. Because the angular momentum is a vector, if the 
velocity vectors of each annulus are not aligned, as is the case for warps, 
then 
the amplitude of the total angular momentum is smaller than the sum of the 
angular momenta in each annuli. We do not correct for this effect in our 
analysis. The fitted inclination warps are of the order of 10\,deg ore less. 
This would translate into a 2\,per\,cent decrease in angular momentum. 

Note, this way to estimate a ``stellar'' specific angular momentum assumes, 
that the kinematics of the stellar disc behave as the kinematics of the \hi\ 
disc. In particular in the centres of galaxies, this is not always the case 
\citep{Bershady2010,Cortese2016a,Cortese2014}

The  baryonic specific angular momentum ($j_{\rm B}$), together with the 
velocity dispersion ($\sigma$), as fitted with \tirific, and the baryonic mass 
($M_{\rm B}$) determine the global stability parameter $q$ 
\citep{Obreschkow2016}:
\begin{equation}
\label{equ:q}
    q = \frac{j_{\rm B} \cdot \sigma}{G \cdot M_{\rm B}}
\end{equation}
(G is the gravitational constant). The values measured as described in this 
section will be used to analyse the \hix\ galaxies in the next section.

When attempting to estimate the dark-matter halo spin parameter for real 
galaxies, we have to make some assumptions. Here we use equ. (19) of 
\citet{Obreschkow2014} and follow their suggested values for the ratio of 
cold baryon mass (i.\,e. the sum of the stellar, \hi\ and H$_2$ masses) to halo 
mass ($f_{\rm M} = 0.05$) and ratio of the specific angular momentum of cold 
baryons to that of the halo ($f_{\rm j} = 1$), based on the results of 
\citet{Stewart2013}, \citet{McMillan2011}, \citet{Flynn2006}, 
\citet{Kalberla2008}, and \citet{Sanders1984}. This leads to the following 
equation:
\begin{equation}
\label{equ:lambda_pragmatic}
    \lambda = \left[ \frac{M_{\rm B}{\rm [M_{\odot}]}}{10^{10}} \right]^{-2/3} 
\times 
\frac{j_{\rm B}{\rm [kpc~km~s^{-1}]}}{10^3} \times 0.069
\end{equation}
Note, however, that in practice, there is a large scatter in the quantities 
that are used in this calculation
\citep{Romanowsky2012,Obreschkow2014,Stevens2016,Lagos2017a}.

\subsection{Data for comparison with HIghMass galaxies}
\label{sec:highmass}
\begin{table*}
\begin{tabular}[H]{l || c c c c }
~& UGC\,9037 & UGC\,12506 & UGC\,6168 & UGC\,7899 \\
$\lambda^{*}$ & 0.07 & 0.15 & 0.09 & 0.08 \\
$\rm \log\,M_{HI}\,[M_{\odot}]^{*}$ & 10.33 & 10.53 & 10.35 & 10.42 \\
$\rm \log\,M_{\star}\,[M_{\odot}]$  & 10.45 & 10.79 & 10.43 & 10.82 \\
$\rm \log\,M_{B}\,[M_{\odot}]$      & 10.75 & 11.03 & 10.76 & 11.01 \\
j$_{B}\,{\rm [kpc~km~s^{-1}]}$ & 3239 & 10583 & 4169 & 5438 \\
$\rm \log\,q$ & -0.84 & -0.60 & -0.73 & -0.86 \\
\end{tabular}                                                                  
    \caption{Properties for four \highmass\ galaxies. Properties marked with 
$^{*}$ are taken from \citet{Hallenbeck2014,Hallenbeck2016}. All other 
properties are calculated as described in the text. }
\label{tab:highmass}
\end{table*}

We will compare the specific angular momentum and halo spin properties of \hix\ 
galaxies to four \highmass\ galaxies \citep{Huang2014}, for which resolved \hi\ 
maps and accurately measured halo spin parameter are published 
\citep{Hallenbeck2014,Hallenbeck2016}. The measurements for these galaxies are 
summarise in Tab.\ref{tab:highmass}. The stellar mass of these galaxies has 
been calculated from \kband\ photometry the same way as for \hix\ galaxies 
(see Equ. \ref{equ:mstar}). The baryonic mass is calculated from stellar and 
\hi\ mass as detailed in Equ. \ref{equ:mbary}. The baryonic specific angular 
momentum has been calculated from the halo spin parameter using Equ. 
\ref{equ:lambda_pragmatic}. The stability parameter is then determined as for 
the \hix\ galaxies (see Equ. \ref{equ:q}), assuming a velocity dispersion of 
$\sigma=11\,{\rm km\,s^{-1}}$. 

There has been a fifth \highmass\ galaxy (NGC\,5230) for which resolved \hi\ 
maps have been published, however, due to its low inclination, 
\citet{Hallenbeck2016} refrain from measuring a spin parameter in this galaxy.

\section{Results}
\label{sec:results}
\subsection{Masses and sizes}

\begin{figure}
\includegraphics[width=3.15in]{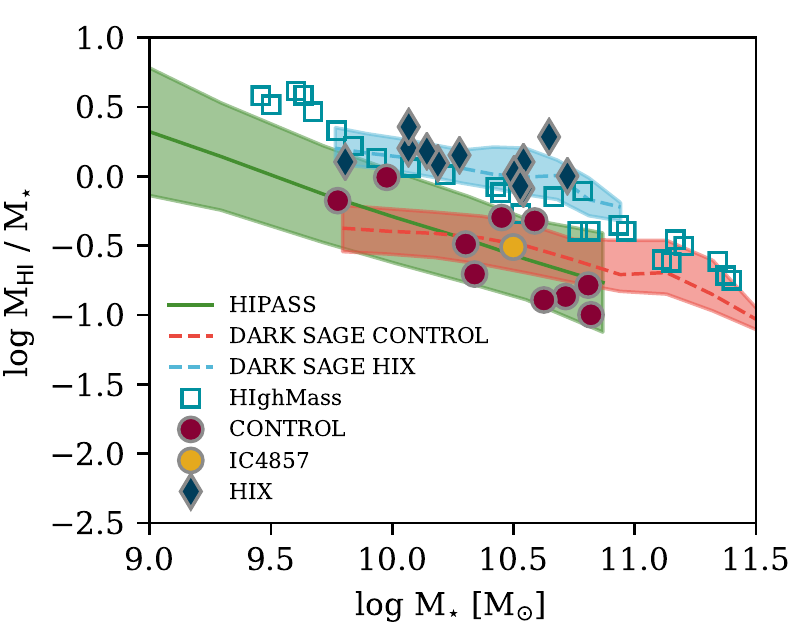}
\caption{The \hi--to--stellar mass ratio as a function of the stellar mass. 
The green shaded area shows the $1\,\sigma$ range of the parent sample (with 
\mass\ cross match), 
red circles give the \control\ sample and blue diamonds represent the \hix\ 
sample. The yellow circle is IC\,4857, which was initially selected as a \hix\ 
galaxy, but then reclassified to a \control\ galaxy. The empty blue squares 
present the \highmass\ sample. Orange and light blue dashed lines indicate the 
running average of simulated \hix\ (light blue) and \control\ galaxies (orange) 
from the \ds\ semi-analytic model (for more details see Sec. 
\ref{sec:darksage}). The orange and light blue shading covers the 16 to 84 
percentile range. As per sample selection, \hix\ galaxies have high \hi\ mass 
fractions for their stellar mass.}
\label{fig:mhi_vs_mstar}
\end{figure}

As detailed in \citet{Lutz2017}, the \hix\ galaxy sample was selected to 
contain more \hi\ than expected from their stellar luminosity using 
scaling relations by \citet{Denes2014}. This translates into \hix\ 
galaxies containing a high \hi\ to stellar mass fraction for their stellar 
mass. Fig. \ref{fig:mhi_vs_mstar} shows that this the case for most \hix\ 
galaxies. So far we have found that the stellar discs of the \hix\ galaxies are 
similar to those of the \control\ sample in terms of star formation activity 
and morphology. This means that the extreme \hi\ content of \hix\ galaxies does 
not drive extreme star formation.

\begin{figure}
\includegraphics[width=3.15in]{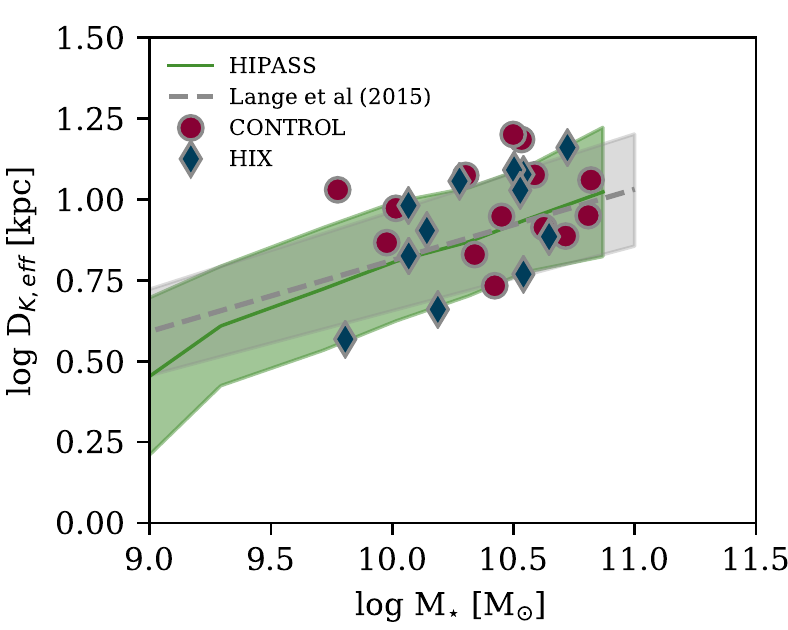}
\caption{A stellar mass--size relation. The running average of the \hipass\ 
parent sample is shown as the solid, green line, with the $\pm 1\,\sigma$ 
scatter as the green shaded area. For comparison the grey dashed line and grey 
shaded area give the mass--size relation and its errors for late-type galaxies 
from the GAMA survey \citep{Lange2015}. The \control\ and  \hix\ sample 
galaxies are shown as red circles and blue diamonds respectively. The \hix\ and 
\control\ sample galaxies occupy a similar parameter space and follow the 
relation from the literature and the parent sample.}
\label{fig:reff_vs_mstar}
\end{figure}

We now investigate a stellar mass--size relation. The relation 
between stellar mass and the \kband\ effective radius is shown in Fig. 
\ref{fig:reff_vs_mstar}. As stellar masses are calculated from \kband\ 
luminosities, radii in the \kband\ are treated as equivalent to 
stellar radii. For the galaxies in the \hipass\ parent sample that are 
identified with a 2MASX counterpart, a 
correlation between $M_{\star}$ and $R_{\star}$ is seen. This 
relation is in good agreement with the corresponding scaling relation for 
GAMA galaxies \citep{Lange2015}. Both the \hix\ 
and the \control\ sample follow this relation. Thus in terms of average stellar 
surface density, the stellar discs of \hix\ galaxies are similar to the 
stellar disc of the \control\ sample. 

\begin{figure}
\includegraphics[width=3.15in]{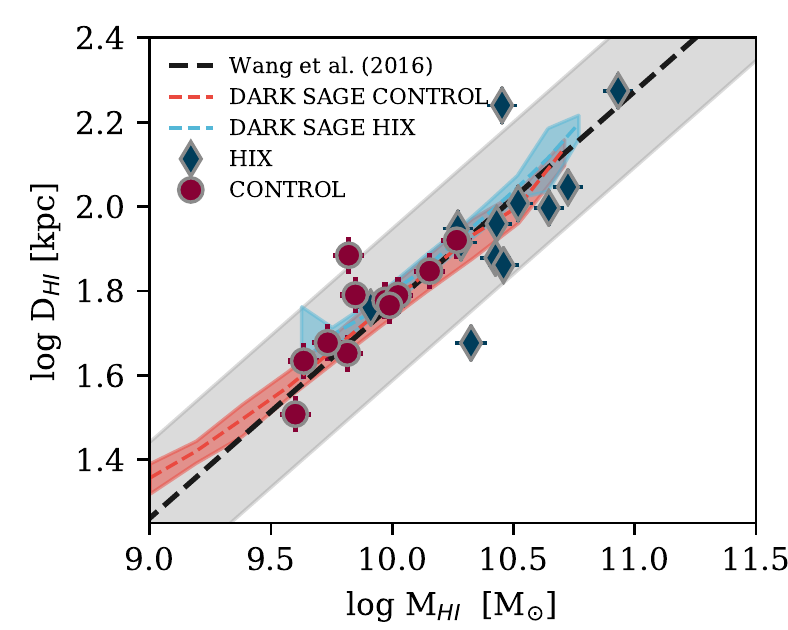}
\caption{The relation between \hi\ disc size and mass. Blue diamonds present 
the \hix\ sample and red circles the \control\ sample. The grey dashed line is 
the relation found by \citet{Broeils1997}, confirmed by \citet{Wang2016}, where 
the grey shaded area covers their $3\,\sigma$ scatter of 0.18 dex. As in Fig. 
\ref{fig:mhi_vs_mstar} light blue and orange dashed lines present \ds\ 
simulated galaxies and the shaded areas their respective 16 to 84 percentile 
ranges. \hi\ masses and sizes of the \hix\ galaxies are consistent with the 
literature relation.}
\label{fig:dhi_vs_mhi}
\end{figure}

Fig. \ref{fig:dhi_vs_mhi} shows the relation between the \hi\ disc size and 
\hi\ mass for the \hix\ and the \control\ sample. Both samples follow the 
\citet{Wang2016} relation, which was fitted to around 400\,galaxies and is in 
good agreement with the previous relation by \citet{Broeils1997}. All \hix\ and 
\control\ galaxies have average \hi\ column densities and their \hi\ discs 
behave as average \hi\ discs in the local Universe.

\begin{figure}
\includegraphics[width=3.15in]{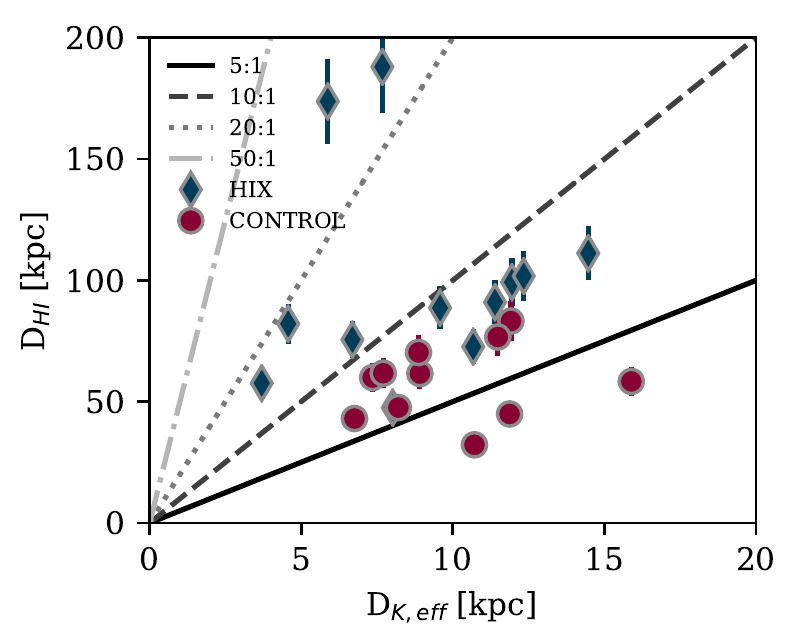}
\caption{The relation between $R_{\rm HI}$ and the \kband\ effective radius. 
Blue diamonds present the \hix\ sample and red circles the 
\control\ sample. Grey lines denote different ratios between the two sizes. 
At a given \kband\ effective radius, \hix\ galaxies tend to have larger \hi\ 
disc sizes than the \control\ sample. }
\label{fig:dhi_vs_d25}
\end{figure}

Both the \hi\ and the stellar discs of the \hix\ and \control\ samples fall on 
the same respective mass--size relations. However, the \hix\ galaxies have more 
massive \hi\ discs than the \control\ sample. Therefore, we also compare the 
$R_{\rm HI}$ and the \kband\ effective radius in Fig. \ref{fig:dhi_vs_d25}. As 
expected, the \hi--to--stellar disc size ratio is larger for most \hix\ 
galaxies. Exceptions are two \hix\ galaxies, for which the data points are 
located within the scatter of the \control\ sample data. 
These two galaxies are ESO461-G010 and ESO290-G035. 

ESO461-G010 and ESO290-G035 are the two only \hix\ galaxies with \tirific\ 
inclination angles above 80\,deg. This might hamper the measurements. 
There are also four control galaxies with such high inclinations, these 
galaxies are however at smaller distances and thus better resolved. 

The two galaxies with the largest \hi\ to stellar diameter ratios and \hi\ 
diameters above 150\,kpc are NGC\,289 and ESO075-G006. The ``smaller'' of the 
two galaxies is NGC\,289. This galaxy has been studied extensively before.  
\citep{Walsh1997} has classified NGC\,289 as a low surface brightness galaxy 
with a high surface brightness inner disc. Latest photometry shows a central 
surface brightness in \rband\ of $\sim17$\magasec \citep{Li2011}, i.\,e. the 
central disc is a high surface brightness disc\footnote{ for detailed 
photometric profiles see 
https://cgs.obs.carnegiescience.edu/CGS/object\_html\_pages/NGC289.html}. In 
addition, NGC\,289 has been found to host an extended UV disc (XUV disc) and 
has star forming regions beyond the optical disc \citep{Meurer2017}. There is a 
 potential dwarf galaxy, PGC\,708504, which is not detected by \mass\ and is 
not detected as a separate \hi\ source in our ATCA data. In particular 
in that respect NGC\,289 is remarkably similar to ESO075-G006, which also has a 
nearby dwarf companion without an \hi\ detection (for a more 
detailed discussion see \citealp{Lutz2017}). The \galex\ \textit{NUV} image of 
ESO075-G006 has a much shorter exposure time than the one of NGC\,289 (207\,s 
compared to 1696\,sec) and optical imaging is of lower quality. Hence, a XUV 
disc or an extend low surface brightness stellar disc can not be excluded in 
ESO075-G006. However, given the large \hi\ column densities above 1\msunpcsq\ 
well outside the visible disc of ESO075-G006, star formation in an XUV disc is 
also a possibility in this galaxy. If this is the case, then galaxies with very 
large \hi\ to stellar disc sizes might be galaxies with a high surface 
brightness inner disc and an extended low surface brightness outer disc. 

Interestingly, the \hi\ disc of ESO208-G026 is approximately 7.5\,times larger 
than the stellar disc, despite the fact that this galaxy is located less than  
$1\,\sigma$ above the running average of \hipass\ in Fig. 
\ref{fig:mhi_vs_mstar}. In the \hi\ mass--size plane, this galaxies lies right 
on the \citet{Wang2016} relation. In the stellar mass--size plane, this galaxy 
is located approximately $1\,\sigma$ below the running average of \hipass. 
ESO208-G026 has been classified as S0 by the RC.3 catalogue 
\citep{deVaucouleurs1991} and upon inspection of optical images, a bright bulge 
and a very faint disc can be seen (see App. \ref{app:hix}). Therefore the 
effective radius is small. There is a second S0 galaxy in the \hix\ sample: 
ESO055-G013. This galaxy has a gas mass fraction more than $1\,\sigma$ above 
the \hipass\ running average. In terms of stellar and \hi\ radii it still 
behaves similar to ESO208-G026. 

\subsection{Warps and tails in HI discs}
\label{sec:morph}
In this section we compare the grade of lopsidedness, warps and tails in the 
\hi\ discs of the \control\ and \hix\ galaxies. We use the results from the 
\tirific\ modelling and visually inspect the moment maps of both samples. 

One out of twelve \hix\ (8\,per\,cent) and \change{three out of eleven 
\control\ galaxies (27\,per\,cent) }are well fitted with a FLAT disc model. The 
remaining galaxies are fitted with a WARP model. Hence, a higher percentage of 
\control\ than \hix\ galaxies can be described as flat discs. 

\begin{figure}
    \includegraphics[width=3.15in]{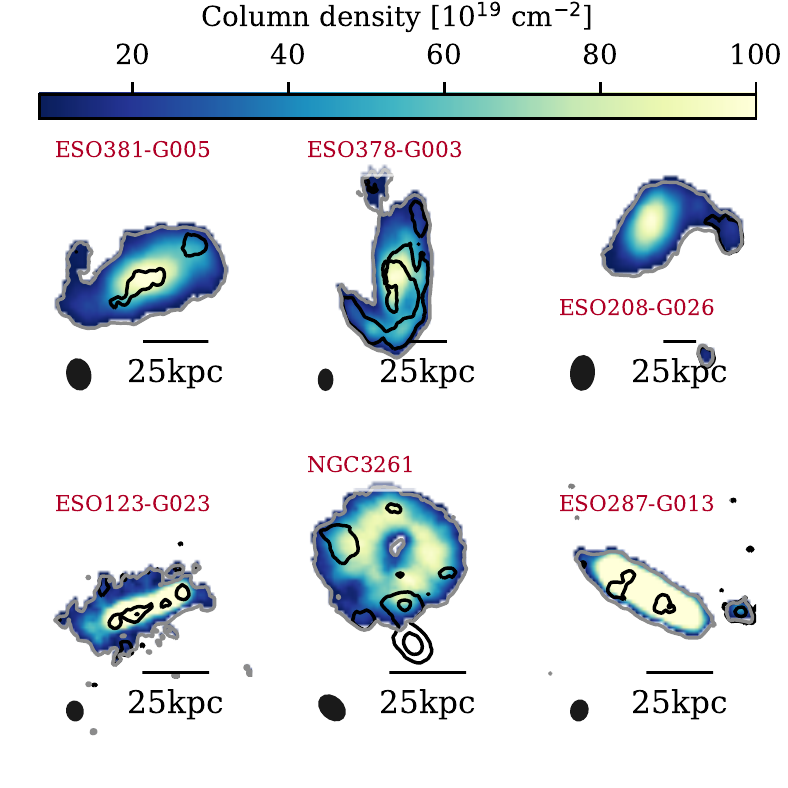}
    \caption{Examples for \hi\ moment 0 maps for \hix\ (top row, from left to 
right: ESO208-G026, ESO378-G003 and ESO381-G005) and \control\ galaxies (bottom 
row, from left to right: ESO123-G023, NGC3261 and ESO287-G013). The colour 
scale indicate \hi\ column densities of the measured data, in the range 
from 8 to $\rm100\times10^{19}\,cm^{-2}$. The grey contour shows the lower 
limit of the colour scale at $\rm8\times10^{19}\,cm^{-2}$. The overlaid black 
contours indicate structures in residual moment 0 maps also at column densities 
of $\rm8\times10^{19}\,cm^{-2}$. Red crosses mark locations of nearby dwarfs 
or potential stellar debris from minor mergers. The images are 3.6\,arcmin in 
size, except for the image of ESO378-G003, which is 6\,arcmin in size. \update}
\label{fig:morph}
\end{figure}

If a disc is fitted as a warped disc, the range of the inclination is of the 
same magnitude in both the \hix\ and the \control\ galaxies. The span of the 
position angles in one fit tends to be larger in \hix\ than \control\ galaxies. 
This trend is mostly driven by \hix\ galaxies ESO208-G026, ESO378-G003 and 
ESO381-G005. These galaxies host prominent arm features (see Fig. 
\ref{fig:morph} and App. \ref{app:hix}), which \tirific\ models with a large 
warp in position angle. We will further discuss these three galaxies below. 

The thickness of the disc is also modelled in \tirific. \change{However, the 
disc thickness is not resolved in most cases. This means that measurements 
of disc thickness can only be upper limits. Within these limitations, }\hix\ 
galaxy models have thicker discs than \control\ galaxies. However, when 
comparing the ratio between disc thickness and disc size, both samples are 
again more similar, with the ratio being around 5 to 10\,per\,cent in most 
cases. \change{The ratio between disc size and thickness is more meaningful 
to decide between thick and thin disc because at a given absolute disc 
thickness only this ratio can inform whether a thin disc or a spheroid is at 
hand. }Outliers with larger ratios are \hix\ galaxies ESO243-G002, ESO055-G013 
and ESO075-G006. Following the Galactic Fountain model, thick discs might 
indicate active gas accretion in these galaxies. \change{However, these three 
galaxies are among the four galaxies farthest away. Thus, their disc thickness 
might be most overestimated by beam smearing.}

% In App. \ref{app:hix} and \ref{app:con}, we show the data of the \hix\ and 
% \control\ sample respectively. In panels (a) and (c), the moment 0 and 1 maps 
% for each galaxy are shown. As discussed above, \hix\ galaxies tend to be less 
% regular with respect to shape and kinematics than \control\ galaxies. For 
% each 
% galaxy, panels (b) show the azimuthally averaged radial \hi\ column density 
% profiles. The average profile for the \control\ and for the \hix\ sample 
% appear 
% similar when the radius is normalised to $R_{HI}$. 
% In panels (e), spectra form the ATCA data and the residual of the \tirific\ 
% model  are shown. On average, we recover 93\,per\,cent and 
% 90\,percent of the \hipass\ \hi\ mass in the ATCA \hi\ data cubes of \hix\ 
% and 
% \control\ galaxies respectively. The bottom two panels (g) and (h) show 
% position--velocity diagrams along the minor and major axis respectively. 
% These 
% diagrams are a good tool to asses the quality of the \tirific\ fit and to 
% search for extra planar gas and a thick \hi\ disc.  

As mentioned above, galaxies ESO208-G026, ESO378-G003 and ESO381-G005 are 
fitted with the strongest warps. Upon visual inspection, these galaxies also 
show the most interesting \hi\ morphologies with \hi\ tails, arms and clouds. 
Images in the top panel of Fig. \ref{fig:morph} show the moment 
0 maps of these galaxies. To further quantify the contribution of these 
peculiar morphologies to \hi-richness, we generate (not fit) flat disc model 
data cubes for all galaxies with \tirific. These cubes are generated with the 
following parameters:
\begin{itemize}
    \item the previously fitted \tirific\ column density profile, 
    \item the fitted rotation curve (functional form as in Equ. \ref{equ:vrot}),
    \item radially constant position angle and inclination (the median of the 
warped profiles), 
    \item velocity dispersion, systemic velocity and kinematic centre as found 
previously with the fully fitted \tirific\ model. 
\end{itemize}
For galaxies with a FLAT \tirific\ fit, these generated cubes are the same as 
the fitted model cube. For galaxies with a warped \tirific\ fit, these 
generated data cubes are ``rectified'' versions of the fitted model cube. 

In a next step we subtract the generated cube from the observed data cube. If 
the moment 0 map of this residual cube shows structure, the \hi\ mass is 
measured. This is the case for $(42\pm19)$ and $(36\pm18)$\,per\,cent of the 
\hix\ and the \control\ galaxies, respectively. Hence, within Poisson errors, 
residual structures are found in the same percentage of \hix\ galaxies as 
\control\ galaxies. Fig. \ref{fig:morph} shows examples for \hix\ (top row) and 
\control\ galaxies (bottom row) with residual structures (black contours). 

For those galaxies with residual structure, the average fraction of residual 
\hi\ to total \hi\ mass is 
change{$(19\pm6)$ and $(9\pm8)$\,per\,cent }for \hix\ and 
\control\ galaxies. \change{Hence, \hix\ galaxies tend to host more \hi\ }that 
does not agree with a flat, regularly rotating disc than \control\ galaxies. 
\change{It has to be noted though that the residuals of the full model data 
cubes are of the same order. }However, according to this measurement, the 
irregular \hi\ can not fully account 
for the higher than expected \hi\ mass in \hix\ galaxies. So irregular gas is 
not the sole driver of \hi\ excess in \hix\ galaxies. In 
addition some of the irregular gas in \control\ galaxies might be missed due to 
the less sensitive observations. 

The reason for the irregularity of the \hix\ galaxies in Fig. \ref{fig:morph} 
can only be speculated about:
ESO208-G026 is located in a relatively sparse environment. Hence, it would be 
more likely for the arm to be caused by a gas-rich minor merger than by tidal 
interaction. However, the stellar disc of  ESO208-G026 appears regular and no 
debris of the merger is found in optical images. 

ESO378-G003 is part of the NGC\,3783 group \citep{Kilborn2006} and the 
elliptical galaxy NGC\,3706 is located 22\,arcmin = 255\,kpc away (projected 
distance). Furthermore, 
\citet{Kilborn2006} has found an isolated \hi\ cloud (GEMS\_N3783\_2) 
without apparent optical counter part (see their fig. 6 for a resolved map). 
They concluded the cloud to be a remnant of tidal interaction. This tidal 
interaction could have potentially disturbed the \hi\ distribution of 
ESO378-G003 as well. 

In ESO381-G005 the arm feature is aligned with wide spiral arms 
and points towards a dwarf companion (PGC\,629239) towards the south, which is 
also detected in \hi. Additionally the inspection of optical images reveals 
three more sources aligned within the arm that are not foreground stars 
(PGC630365, 
USNO~A2~0525-15319577 and USNO~A2~0525-15322378), but look more like 
potential debris. Their position on the \hi\ 
column density map is marked with black crosses in Fig. \ref{fig:morph}. This 
might indicate a recent minor merger. However, in order for the minor merger to 
increase the \hi\ content of a galaxy from \control\ sample levels to \hix\ 
sample levels, it would have had to bring in almost $\rm 10^{10}\,M_{\odot}$ 
without changing the stellar disc a lot. This appears very unlikely to be the 
only cause why ESO381-G005 is a \hix\ galaxy.

In summary, both the \hix\ and \control\ galaxies show warped discs. \hix\ 
galaxies tend to be fitted with stronger warps and are more likely than the 
\control\ galaxies to have tails or ill-defined outskirts. As mentioned 
before, the differences in sensitivity of the observations might increase the 
difference in \hi\ morphology between \hix\ and \control\ galaxies. 
Homogeneous, large surveys of the resolved \hi\ content of galaxies, 
such as the Wallaby survey, will help to shed more light onto this 
problem. Even in the most irregular \hix\ galaxies, the amount of 
\hi\ in arms and tails appears not to be sufficient to explain why these 
galaxies host more than 2.5 times more \hi\ than expected. 

\subsection{Global stability parameter q}
\label{sec:stabel}

So far the analysis has shown that the most striking difference between \hix\ 
and \control\ sample is the difference in the \hi\ disc mass and subsequently 
size. It has long been known that the galaxy size is determined by 
the angular momentum of the galaxy \citep{Fall1980,Dalcanton1997,Mo1998}. More 
recently, \citet{Maddox2015} and \citet{Obreschkow2016} have suggested 
based on \alfalfa\ \citep{Giovanelli2005} and THINGS data \citep{Walter2008} 
that the \hi\ mass (and thus size), and atomic--to--baryonic mass fraction is 
regulated by the angular momentum properties of the galaxy. \citet{Huang2014} 
and \citet{Hallenbeck2014} suggest that the \hi-rich galaxies in the 
\highmass\ sample are \hi-rich due to to an increased specific angular 
momentum. A higher specific angular momentum affects the disc in two ways:
\begin{enumerate}
    \item Gas is stabilised against gravitational collapse and subsequent star 
        formation.
    \item \hi\ is kept at larger galactocentric radii, where the total disc 
        density is too low to form molecular hydrogen and stars. 
\end{enumerate}
Thus, a galaxy with a higher specific angular momentum can support a larger 
\hi\ disc. This stability can be quantified as the 
global stability parameter $q$ (\citealp{Obreschkow2016}, see also Sec. 
\ref{sec:radprof}), which is proportional to a disc wide average of the local 
Toomre $Q$ parameter \citep{Toomre1964}. A more descriptive way to 
understand and interpret this parameter is the following: Consider the radial 
variation of the Toomre $Q$ parameter in two galaxies with a similar 
exponential discs (i.\,e. similar scale radii) but different $q$'s. In the 
galaxy with the larger $q$, \hi\ begins to be Toomre-stable at smaller radii 
than in the galaxy with a smaller q (see fig. 1 in \citealp{Obreschkow2016}). 

\begin{figure}
\includegraphics[width=3.15in]{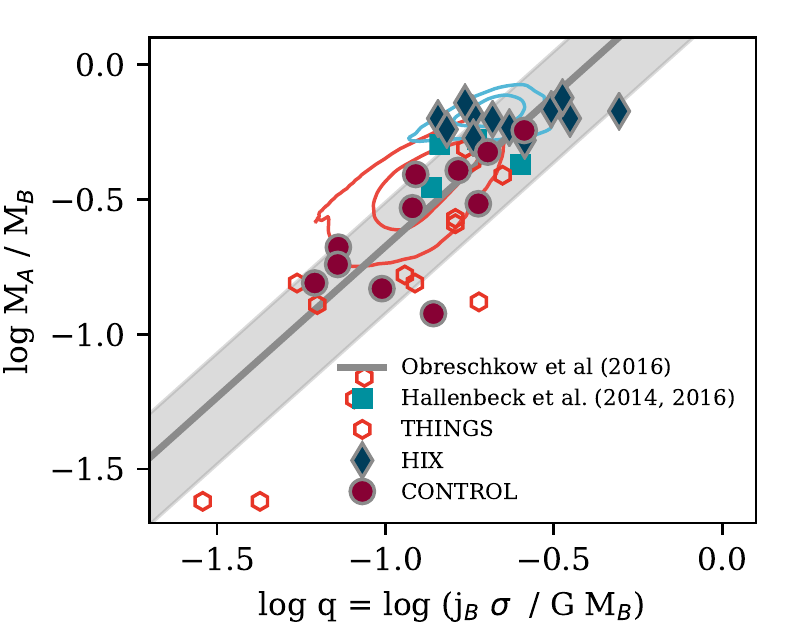}
\caption{The atomic to baryonic gas ratio as a function of the global stability 
parameter. In addition to the data of the \hix\ (blue diamonds) and \control\ 
sample (red circles), we also show data of THINGS galaxies (orange pentagons, 
\citealp{Obreschkow2014}) and \highmass\ galaxies (light blue squares). The 
orange and light blue contours encompass 68 and 95\,per\,cent of the simulated 
\control\ and \hix\ galaxies from \ds\ (for more details see Sec. 
\ref{sec:darksage}). The data of observed and simulated galaxies agree with the 
analytical model of \citet{Obreschkow2016}. }
\label{fig:fatom_vs_q}
\end{figure}

Fig. \ref{fig:fatom_vs_q} shows the atomic--to--baryonic mass fraction as a 
function of $q$. Most of the \hix\ and the \control\ sample follow this 
model within the scatter of the analytical model of \citet{Obreschkow2016}. 
This means that galaxies from both samples host as much \hi\ as 
they can support against star formation. For comparison, we also show THINGS 
galaxies (data taken from \citealp{Obreschkow2014}) and \highmass\ galaxies 
\citep{Huang2014,Hallenbeck2014,Hallenbeck2016} (see Sec. \ref{sec:highmass} 
for more details on the data). Both samples also follow the model.

Most \hix\ galaxies have systematically larger global stability $q$ values than 
the \control\ galaxies. This is due to a larger baryonic specific angular 
momentum, which is driven by a larger \hi\ specific angular momentum. Rotation 
velocities are of similar magnitude in \hix\ and \control\ galaxies (see Tab. 
\ref{tab:props}). So the large \hi\ specific angular momenta in \hix\ 
galaxies are mostly due to large masses of high angular momentum gas located at 
large galactocentric radii. 

The question now is: why do \hix\ galaxies have high specific \hi\ angular 
momenta. In the cases of galaxies, ESO378-G003, ESO381-G003 and ESO208-G026, 
which were discussed in the previous section, the gas might have been recently 
accreted or dislocated in some interaction. Other galaxies might be living in 
high-spin haloes, which gives them an intrinsically high specific angular 
momentum. To test this hypothesis, we compare our galaxies to simulated 
galaxies from the semi-analytic model \ds\ in the next section. 

\section{Comparison with D{\small ARK} S{\small AGE}}
\label{sec:darksage}
We use the semi-analytic model \ds\ \citep*{Stevens2016} to investigate how the 
dark-matter halo spin affects the \hi\ content of \hix\ galaxies.
Semi-analytic models use halo merger trees, typically produced from a 
cosmological $N$-body simulation, as input for a series of coupled differential 
equations that describe the evolution of galaxies \citep[see reviews 
by][]{Baugh2006,Somerville2015}. \ds\ is an ideal model for our investigation, 
as, other than when a galaxy is first initialised in the model, there are no 
assumptions made about the relationship between the instantaneous spin of a 
(sub)halo and the galaxy it hosts (see \citealt{Stevens2017} and references 
therein for a discussion on the importance of this). Instead, a galaxy's 
specific angular momentum depends on the entire history of its halo's spin 
evolution, along with its merger history, and its own secular evolution.  
Furthermore, in addition to other properties, the model is calibrated to the 
observed \hi\ mass function \citep{Zwaan2005} and to the mean \hi--to--stellar 
mass fraction of galaxies as a function of stellar mass \citep{Brown2015}.

The unitless spin parameter of a dark-matter halo is defined as
\begin{equation}
    \lambda \equiv \frac{J |E|^{1/2}}{G M_{\rm vir}^{5/2}} = 
    \frac{j_{\rm halo}}{\sqrt{2} R_{\rm vir} V_{\rm vir}}~
\label{equ:lambda}
\end{equation}
\citep{Peebles1969,Bullock2001}, where $J$ is the halo's total angular 
momentum, $E$ its total energy, $M_{\rm vir}$ its virial mass, $j_{\rm halo}$ 
its specific angular momentum, $V_{\rm vir}$ its virial velocity, and $R_{\rm 
vir}$ its virial radius.
Simulations show that the spin parameters of haloes follow an approximately 
log-normal distribution with a peak around 0.03 
\citep{Barnes1987,Bullock2001,Shaw2006}. 

At each time-step in {\sc Dark Sage}, gas can cool onto a galaxy from its halo. 
The net specific angular momentum of the gas that cools is assumed to be the  
same as that of halo at that instant (related to $\lambda$ through 
equ.~\ref{equ:lambda}). This gas is distributed with an exponential profile 
into 
30 disc annuli, each of which has a fixed specific angular momentum. The 
angular momentum vector of this gas is then summed with that already in the 
galaxy to define the gas disc's new plane, and both distributions are projected 
onto a new set of annuli in that plane. Frequent changes to the halo's spin 
direction (or magnitude) and mergers can lead to a reduction in the galaxy's 
specific angular momentum. Full details are provided in \citet{Stevens2016}.

\subsection{The simulated galaxy catalogue}
\label{sec:sim_gal}
We obtain a box-type catalogue of simulated galaxies from the Theoretical 
Astrophysical 
Observatory (TAO\footnote{https://tao.asvo.org.au}, \citealp{Bernyk2016}).
The underlying dark--matter--only simulation is the Millennium 
simulation \citep{Springel2005}. We use the full box size of 
$\rm (500\,Mpc/h)^{3}$ and consider the snapshot at $z\!=\!0$. The Millennium 
simulation uses the cosmological parameters from \textit{WMAP-1} 
\citep{Spergel2003}. 
Note, however, that we apply H$_0$ = 70\,km\,Mpc$^{-1}$\,s$^{-1}$ where 
relevant to these data to be consistent with the presentation of our 
observational data.
The dark-matter haloes are populated using the 2016 version of the \ds\ 
semi-analytic model \citep{Stevens2016}. Because \ds\ evolves galaxy discs
in annuli of constant specific angular momentum, the model can make 
predictions on structural and kinematic properties of the 
simulated galaxies. 

In addition to the galaxy parameters that \ds\ computes, TAO 
can also fit the spectral energy distribution (SED) of a galaxy and thus 
calculate its brightness in given filters. In our case, we chose    
\citet{Bruzual2003} SEDs with a \citet{Chabrier2003} initial mass function 
(foreground dust modelling was not included). 
The final catalogue of simulated galaxies includes masses and angular 
momenta for each of stars, \hi\ and molecular gas, as well as bulge--to--total 
mass ratios, \hi\ and stellar disc sizes, absolute \rband\ magnitudes, and 
dark-matter halo spin parameters.

\subsection{Selecting D{\small ARK} S{\small AGE} H{\small IX} and control 
galaxies}
The \hix\ and \control\ sample were selected using the \rband\ -- \hi\ mass 
scaling relation by \citet{Denes2014}. We use the same relation to select 
simulated \hix\ and \control\ galaxies from the \ds\ catalogue. We furthermore 
add a stellar mass cut of $\rm 9.7 \le \log M_{HI} [M_{\odot}] \le 11.2$ and a 
bulge to total mass ratio cut $\rm B/T < 0.3$ to select only disc galaxies. 

This results in a catalogue of 288385\,galaxies of which 18416 (6\,per\,cent) 
are \hix\ and 154730 (53\,per\,cent) \control\ sample--like. On the gas mass 
fraction -- stellar mass relation, the \hi\ mass--size relation, and the 
atomic--to--baryonic mass fraction vs. stability parameter plane the 
simulated and real samples behave similarly (see their running averages in 
Figs. \ref{fig:mhi_vs_mstar} and \ref{fig:dhi_vs_mhi} and their contours in 
Fig. \ref{fig:fatom_vs_q}). 

\subsection{Properties of simulated H{\small IX} and \control\ galaxies}
\begin{figure}
    \includegraphics[width=3.15in]{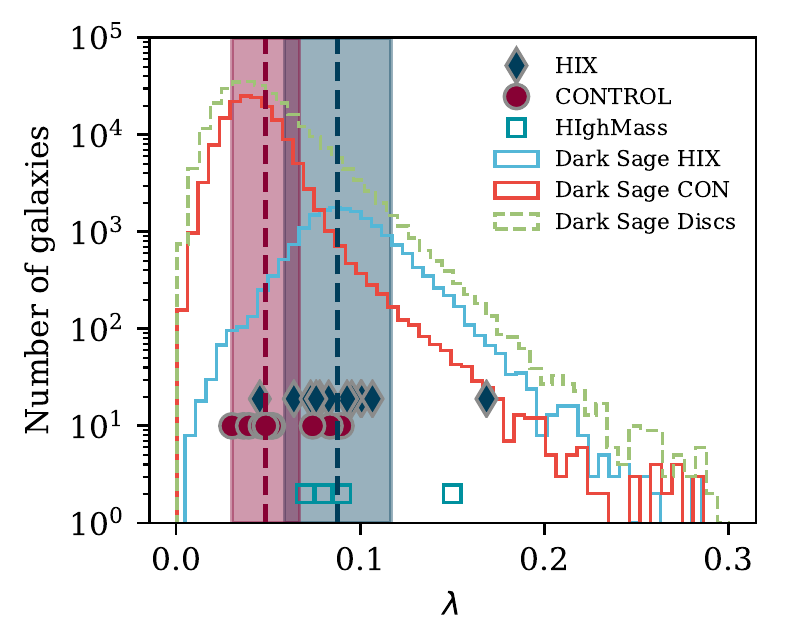}
    \caption{Distributions of the halo spin parameter for the \ds\ model 
galaxies (which come directly from Millennium data). The entire disc sample 
is shown in the green dashed 
histogram, \hix--like galaxies in the light blue and \control--like in 
the orange histogram. Estimates for the halo spin of single \hix\ and  
\control\ galaxies are shown in blue diamonds and red circles and the sample 
medians are marked with the vertical blue and red dashed line, 
respectively. The 1\,$\sigma$ range of the \hix\ and \control\ galaxies is 
shaded in the respective colours. For comparison, the spin parameters for four 
\highmass\ galaxies as taken from \citet{Hallenbeck2014,Hallenbeck2016} are 
given as light blue squares. \hix\ galaxies tend to live in haloes with larger 
spins.}
    \label{fig:lambda}
\end{figure}

In Fig. \ref{fig:lambda} the dark-matter halo spin distribution of the entire 
\ds\ disc sample, the \hix, and \control--like \ds\ galaxies are shown. The 
peak of the halo spin distribution of the \hix--like \ds\ galaxies is 
at much higher values than for the \control--like galaxies, which agree with 
the entire \ds\ sample.

The estimates for the halo spin of the observed \hix\ and \control\ galaxies 
show a similar albeit weaker trend: the median estimated halo spin (dashed 
line) is larger for \hix\ than \control\ galaxies. \change{The galaxy with the 
largest 
spin parameter ($\lambda=0.17$) in the two samples 
is a  \hix\ galaxy: ESO461-G010.} 
% his galaxy is also the \control\ 
% galaxy 
% with the most irregular \hi\ disc. As can be seen on its data panel in App. 
% \ref{app:con} and in the bottom, left image in Fig. \ref{fig:morph}, it has 
% ill-defined and diffuse edges. In the \hi\ mass fraction vs. stellar mass plot, 
% ESO123-G023 is one of the \control\ galaxies at the upper edge of the 
% \hipass\ $1\,\sigma$ scatter. In fact it is the \control\ galaxy with the 
% highest \hi\ mass fraction. 

\change{The \control\ galaxy with the highest spin parameter is 
ESO123-G023 with $\lambda=0.09$. The spin parameters of 8 out of 11 
\control\ galaxies are lower than the spin parameters of 11 (out of 12) \hix\ 
galaxies. Only \hix\ galaxy ESO243-G002 has $\lambda = 0.05$. }

The four \highmass\ galaxies that are shown in Fig. \ref{fig:lambda}, are 
those \highmass\ galaxies with published resolved \hi\ maps. For the fifth 
\highmass\ galaxy with a published resolved \hi\ map no spin parameter has been 
published. The \highmass\ galaxies show mostly spin parameters of the same 
order as the \hix\ galaxies, except for galaxy UGC\,12506 
\citep{Hallenbeck2014}, which lives in a very high-spin halo. This further adds 
to the evidence from \citet{Lutz2017} that \highmass\ galaxies are similar to 
\hix\ galaxies. 

In summary, we were able to select model galaxies from \ds, that behave similar 
to our observed \hix\ and \control\ galaxies. When looking at the halo spin 
parameters of the modelled galaxies, \hix--like galaxies tend to live in higher 
spin haloes than \control--like galaxies. The estimates of the halo spin 
parameter in our observed \hix\ and \control\ galaxies show a similar trend. 

\section{Discussion}
\label{sec:discussion}

\hix\ galaxies are more likely to host warped discs and/ or arms and tails than 
\control\ galaxies. Fewer \hix\ galaxies can be described with 
a flat disc. Some diffuse gas in the \control\ galaxies might be missed 
due to less sensitive observations. Hence fewer warps might be fitted to the 
data of \control\ galaxies. Still, the majority of both the \hix\ \textbf{and} 
the \control\ galaxies show warps. The cause of warps is not yet fully 
understood. They can be for example be created in hydrodynamical, cosmological 
simulations when the angular momenta of the inner disc and of the surrounding 
halo or accreted gas are misaligned \citep{Roskar2010}. Hence, the detection of 
warps in both samples can indicate cold gas accretion. Warps can also be formed 
by minor mergers \citep{Sancisi2008}. Many of our sample galaxies have dwarf 
galaxies nearby, some of which are also detected in \hi. If gas-rich, minor 
mergers can not only cause warps, but can also contribute to the \hi\ content 
of 
the galaxy. Furthermore, especially the \hix\ \hi\ discs reach far into the 
halo. There they might be more susceptible to interactions with sub-haloes 
or torques by a mis-aligned outer halo and less supported by the stellar disc. 
These effects can also lead to warps \citep{Jozsa2007a}. 

There is one \hix\ galaxy that shows clear signs of interaction 
both in its stellar and \hi\ disc (ESO245-G010), and three \hix\ galaxies that 
have some kind of \hi\ arms or clouds attached to their disc 
(ESO208-G026, ESO378-G003 and ESO381-G005). These features might be due to 
the accretion of gas rich dwarf companions(ESO245-G010 and ESO381-G005) or 
tidal interaction with another galaxy (ESO378-G003 with NGC\,3706). We find, 
however, that the \hi\ mass of those features is not large enough to fully 
explain the high \hi\ masses of \hix\ galaxies. It should be noted that the 
kinematic properties of the discs of these galaxies still agree with a model of 
a marginally stable disc. 

There is no \control\ galaxy with any signs of interaction in the stellar disc, 
but one galaxy hosts irregularly shaped gas at large radii (ESO123-G023). 
Three \control\ galaxies are accompanied by dwarf galaxies that are detected in 
our \hi\ data cubes (ESO287-G013, IC\,4857 and NGC\,3261). In summary, \hix\ 
galaxies tend to be more lopsided than \control\ galaxies, however, the mass 
of irregular \hi\ is not high enough to fully explain why they are \hix\ 
galaxies.

The environments of the \hix\ and \control\ galaxies are very diverse. Some 
galaxies are located just a few Mpc (projected) away from centres of clusters: 
\hix\ galaxy ESO243-G002 resides 2.2\,Mpc away from ABELL\,2836 and \control\  
ESO383-G005 galaxy is located 1\,Mpc away from ABELL\,3563. Other galaxies are 
very isolated. The nearest neighbour within $\pm500\,{\rm km\,s^{-1}}$ in 
recession velocity for example of ESO208-G026 according to NED 
\footnote{http://ned.ipac.caltech.edu, The NASA/IPAC Extragalactic Database 
(NED) is operated by the Jet Propulsion Laboratory, California Institute of 
Technology,under contract with the National Aeronautics and Space 
Administration.}, is ESO208-G031 at a  projected distance of 1.7\,Mpc. 

Resolved \hi\ maps have been published for five \highmass\ galaxies 
\citep{Hallenbeck2014,Hallenbeck2016}. None of these galaxies show any signs of 
arms, tails or diffuse and irregular edges to their column density limits of 
$\rm > 8.4 \times 10^{20}\,cm^{-2}$. This limit, however, is about 10 to 
20\,times higher than our column density limits and thus too high to detect 
arms as hosted by ESO208-G026, ESO378-G003 and ESO381-G005.

When considered independently, \hi\ and stellar discs of both samples follow 
the same mass--size relations (Figs. \ref{fig:reff_vs_mstar} and 
\ref{fig:dhi_vs_mhi}). Hence, the average \hi\ and stellar column densities are 
as average for a galaxy in the local Universe in both samples. 

However, the \hi\ discs of the \hix\ galaxies extend further beyond the 
stellar disc than those of \control\ galaxies. \hix\ galaxies host \hi\ at 
larger galactocentric radii than \control\ galaxies. This gas has 
a larger specific angular momentum and can therefore not flow to the central 
parts of the galaxy. Due to the low overall density (no detected stellar disc 
component) at these large galactocentric radii, this gas is furthermore unable 
to collapse and form stars.

The high specific angular momentum of this \hi\ gas at large radii heavily 
contributes to the total specific baryonic angular momentum, which is higher in 
\hix\ than in \control\ galaxies. Evidence points to the specific baryonic 
angular momentum as a primary regulator of \hi\ disc size and mass. This is in 
agreement with the \citet{Obreschkow2016} model (Fig. \ref{fig:fatom_vs_q}) 
and the results of hydrodynamical simulations \citep{Lagos2017}. 

In \hix\ galaxies, the high \hi\ specific angular momentum is actually the sole 
driver of an elevated baryonic specific angular momentum. The stellar
specific angular momenta (based on \hi\ kinematics) are of similar size in the 
\hix\ and \control\ sample. This observation is reflected in the stellar 
mass--size relation, where both samples are located in similar areas. This 
allows parameters like the specific star formation rate (see section 4.2 in 
\citealp{Lutz2017}) or the average stellar surface density (see Fig. 
\ref{fig:reff_vs_mstar}) to be similar between the two samples. 

From the observations it is not possible to determine how these galaxies 
acquired significant amounts of \hi\ at large radii or why their specific 
baryonic angular momentum is elevated. In Sec. \ref{sec:darksage}, we have used 
the semi-analytic galaxy model \ds\ built on the dark-matter Millennium 
simulation. \hix--like galaxies in this simulated galaxy 
catalogue tend to live in dark-matter haloes with a larger spin than 
\control--like galaxies. The estimated dark-matter spins for \hix\ galaxies 
show a similar, but weaker trend: \change{The bulk of the \hix\ galaxies is 
within 
haloes that have higher spin parameters than the bulk of the \control\ 
galaxies. }

So an explanation for most \hix\ galaxies might be that they are 
galaxies in intrinsically high-spin haloes. Thus, their high baryonic specific 
angular momentum is inherited from the halo. The \hix\ galaxies with a 
relatively low halo spin are ESO111-G014, ESO243-G002 \change{and ESO245-G010}. 
The ratio of halo to baryonic specific angular momentum is 
subject to a lot of variation \citep{Ubler2014,Bett2010}. When we estimate the 
halo spin of the \hix\ galaxies, we assume a constant halo to baryonic specific 
angular momentum ration. So on these galaxies, this ratio might have been 
changed during their history. Simulations suggest some mechanisms that can lead 
to an increased specific angular momentum (and thus an extended \hi\ disc).

Based on the cosmological, hydrodynamical Illustris simulation, 
\citet{Genel2015} and \citet{Zjupa2017} suggest that feedback can tamper with 
angular momentum in the sense that strong stellar winds increase and AGN 
feedback decreases the angular momentum of galaxies. Hence, \hix\ galaxies might 
have undergone stronger star bursts in the past that removed more 
low-angular-momentum gas in galactic winds thus increasing the overall specific 
angular momentum. This is in agreement with cosmological zoom-in simulations by 
\citet{Ubler2014}, who find that galaxies simulated with strong stellar feedback 
have higher angular momenta than galaxies with weak stellar feedback. 

\citet{Stewart2011,Stewart2013,Stewart2016} suggest, based 
on cosmological zoom-in simulations, that gas accreted through cold filamentary 
accretion increases the angular momentum of the galaxy disc. In particular 
\citet{Stewart2013} finds that gas accreted in the cold mode forms a so called 
cold flow disc which corotates with the galaxy disc. They suggest that these 
cold flow discs might be observed as extended \hi\ or UV disc today. In their 
picture, \hix\ galaxies might be galaxies which accreted gas 
that was never shock heated. Furthermore, the specific angular momentum 
can be increased in minor and/ or gas-rich mergers \citep{Lagos2017a}. 

The \hix\ galaxies with a relatively low halo spin might, also be 
galaxies in transition. ESO245-G010 for example shows clear 
signs of a recent merger. Large gas masses at large radii might move further 
towards the centre of the galaxy and be used for star formation, once the disc 
settles. 

% \citet{Warren2007} and \citet{Warren2006} investigated a sample of 
% dwarf galaxies with large \hi\ mass to light ratios, which initially appears 
% similar to our sample selection. They found that these galaxies are actually 
% star poor rather than \hi-rich by showing that their sample does agree with 
% the baryonic Tully-Fisher relation but not a stellar Tully-Fisher relation. 
% We have checked this for our sample as well and find that the \hix\ galaxies 
% lie within the scatter of the \citet{Lelli2016} baryonic Tully-Fisher 
% relation 
% and agree with the stellar Tully-Fisher relation of the \control\ galaxies. 
% Hence, \hix\ galaxies are truly \hi-rich rather than star-poor. 

\section{Summary and Conclusions}
\label{sec:conclude}

In this second paper on the \hix\ galaxies, we have analysed the spatially 
resolved \hi\ distribution and kinematics of the \hix\ and the \control\ 
galaxies. Our findings can be summarised in the following three points:
\begin{enumerate}
    \item The stellar and \hi\ discs of the \hix\ and \control\ samples follow 
the same respective mass--size relations. This means that the stellar discs of 
the \hix\ galaxies behave like average stellar discs, and the \hi\ discs of 
\hix\ galaxies are consistent with average \hi\ discs in the local Universe. 
Only the relation between \hi\ and stellar disc within \hix\ galaxies makes 
them outliers to the \citet{Denes2014} scaling relations.

    \item We find the \hi\ discs of galaxies in the \hix\ sample are more 
likely to be warped and irregular than in the \control\ sample. Yet, 
most \control\ galaxies are also warped and/or host irregularly shaped gas at 
the edge of their discs. Warps and irregular features can be a sign for 
gas-rich minor mergers and cold gas accretion in the \hix\ and the \control\ 
sample. Our analysis, however, suggests that the mass of detected irregular 
gas in \hix\ galaxies is not sufficient to explain their excess in \hi.

    \item \hix\ galaxies have a higher \hi\ and thus baryonic specific angular 
momentum than \control\ galaxies. Thus, \hix\ galaxies are 
consistent with the relation between global stability and atomic to 
baryonic mass fraction by \citet{Obreschkow2016}. This suggests that \hix\ 
galaxies host as much \hi\ as they can support with their baryonic specific 
angular momentum. A comparison to the \ds\ semi-analytic model 
\citep{Stevens2016} suggests that the majority of the \hix\ galaxies have an 
elevated baryonic specific angular momentum because they tend to live in higher 
spin haloes than the majority of the \control\ galaxies. Those \hix\ galaxies 
that do not live in intrinsically high spin haloes might have increased their 
spicific angular momentum over their live time through strong stellar feedback 
that expels low angular momentum gas or through the accretion of high angular 
momentum gas. They might also be galaxies in transition, 
like interacting galaxy ESO245-G010, which are likely to use some of their gas 
in the future.

%     \item The increased halo spin in \hix\ galaxies is however not large enough 
% to also affect the stellar disc of \hix\ galaxies. This might be the crucial 
% difference between \hix\ galaxies and giant low surface brightness galaxies 
% like Malin\,1. 
\end{enumerate}

Our results indicate that \hix\ galaxies will continue to be \hix\ galaxies in 
the future unless they are heavily disturbed by e.\,g. interactions with other 
galaxies. We will continue to investigate the \hix\ galaxies by searching for 
further indications for the accretion of metal-poor gas 
(e.\,g. inhomogeneities or gradients in the gas-phase metallicity distribution) 
in optical spectra obtained for the \hix\ sample and by examining their 
environment from wide field optical imaging.

\section*{Acknowledgments}
We would like to thank the anonymous referee for helpful comments that improved 
the paper. 

KL would like to thank Toby Brown, Luca Cortese, Claudia Lagos, Attila Popping 
and Enrico di Theodoro for inspiring discussion and work experience students 
Devonie Lamb and Zac Broeren for their contribution to the exploration of the 
environment of ESO381-G005, ESO378-G003 and ESO208-G026.

This publication makes use of data products from the Two Micron All Sky Survey, 
which is a joint project of the University of Massachusetts and the Infrared 
Processing and Analysis Center/California Institute of Technology, funded by the 
National Aeronautics and Space Administration and the National Science 
Foundation.

TAO is part of the All-Sky Virtual Observatory (ASVO) and is funded and 
supported by Astronomy Australia Limited, Swinburne University of Technology and 
the Australian Government. The latter is provided through the Commonwealth's 
Education Investment Fund and National Collaborative Research Infrastructure 
Strategy, particularly the National eResearch Collaboration Tools and Resources 
(NeCTAR) and the Australian National Data Service Projects. 

The Australia Telescope Compact Array  is part of the Australia Telescope 
National Facility which is funded by the Australian Government for operation as 
a National Facility managed by CSIRO.

This paper includes archived data obtained through the Australia Telescope 
Online Archive (http://atoa.atnf.csiro.au).

The Parkes telescope is part of the Australia Telescope which is funded by the 
Commonwealth of Australia for operation as a National Facility managed by 
CSIRO. 

This research has made use of the NASA/IPAC Extragalactic Database (NED),
which is operated by the Jet Propulsion Laboratory, California Institute of 
Technology, under contract with the National Aeronautics and Space 
Administration.

This research has made use of the VizieR catalogue access tool, CDS, 
Strasbourg, France. The original description of the VizieR service was published 
in A\&AS 143, 23. 

\bibliographystyle{mnras}
\bibliography{hix2_main}
\clearpage

\appendix

\section{\change{ATCA HI data }of the HIX sample}
\label{app:hix}
In this and the following section we present the data of the \hix\ and the 
\control\ sample respectively that has been used in this paper. For each 
galaxy, there are eight panels:

Panel (a): the \change{ATCA integrated \hi\ intensity (moment 0) }map overlaid 
on the optical \textit{B-band} image from 
SuperCOSMOS. The contours are starting at $0.4 \times 10^{20}$ and $0.8 \times 
10^{20}$\,cm$^{-2}$ for the \hix\ and the \control\ sample respectively and 
double 
with every step. The red ellipse in the bottom left corner shows the 
synthesised beam. 

Panel (b): the radial profile of the \hi\ column density as measured from 
elliptical annuli. \change{The vertical solid grey line marks the \hi\ radius 
$R_{HI}$. Horizontal, dashed lines mark 1\msunpcsq, which is the isophote 
for the $R_{HI}$ measurement and 0.4\,cm$^{-2}$, which of the order of }the 
column density limit for \hix\ galaxies.

Panel (c): the \change{ATCA mean velocity field (moment 1) }map. The red 
ellipse in the bottom left corner shows the synthesised beam.

Panel (d): the rotation velocity measured by TiRiFiC (black dots connected by 
a black dashed line) and a fit to that rotation curve (grey solid line) of the 
functional form:
\begin{equation}
\rm v_{rot} (r) = v_{flat} \cdot \left[ 1 - exp\left(\frac{-r}{l_{flat}}\right) 
\right]
\end{equation}

Panel (e): \hi\ spectra as measured from the ATCA data cube (black solid line), 
from the \change{\hipass\ }cube (grey solid line) and the residual between the 
input data cube and the TiRiFiC model cube (grey dashed line). 

Panel (f): the radial variation of the inclination (black dots and dashed line, 
left y-axis) and the position angle (grey dots and dashed line, right y-axis) 
as modelled by TiRiFiC. 

Panel (g): a position velocity diagram along the minor axis. Blue contours and 
grey scale background present the data cube and red contours the TiRiFiC model 
cube. 

Panel (h): a position velocity diagram along the major axis. Colour coding as 
in Panel (g).

\begin{figure*}
\includegraphics[width=5.5in]{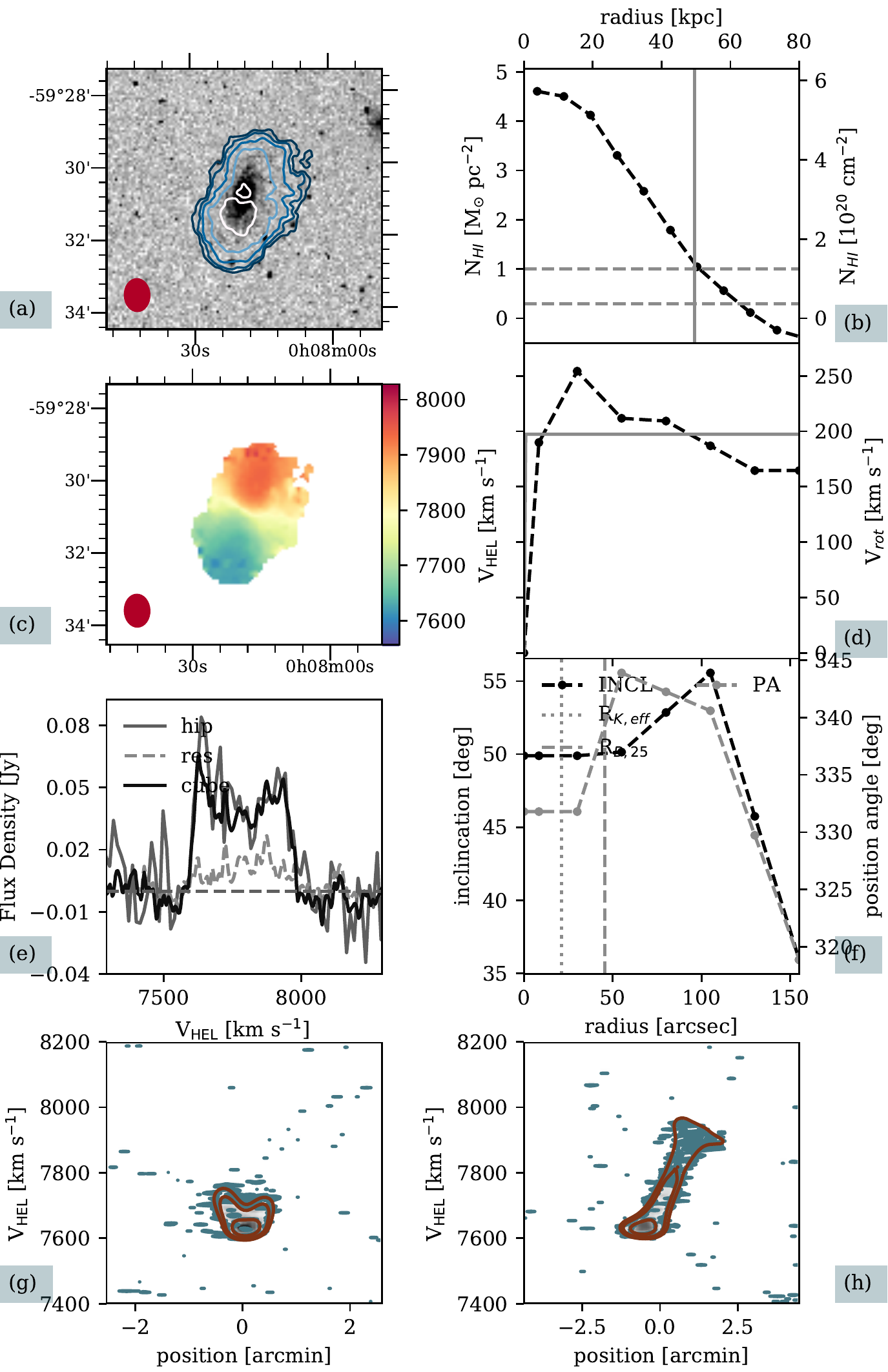}
\caption{ESO111-G014}
\end{figure*}

\begin{figure*}
\includegraphics[width=5.5in]{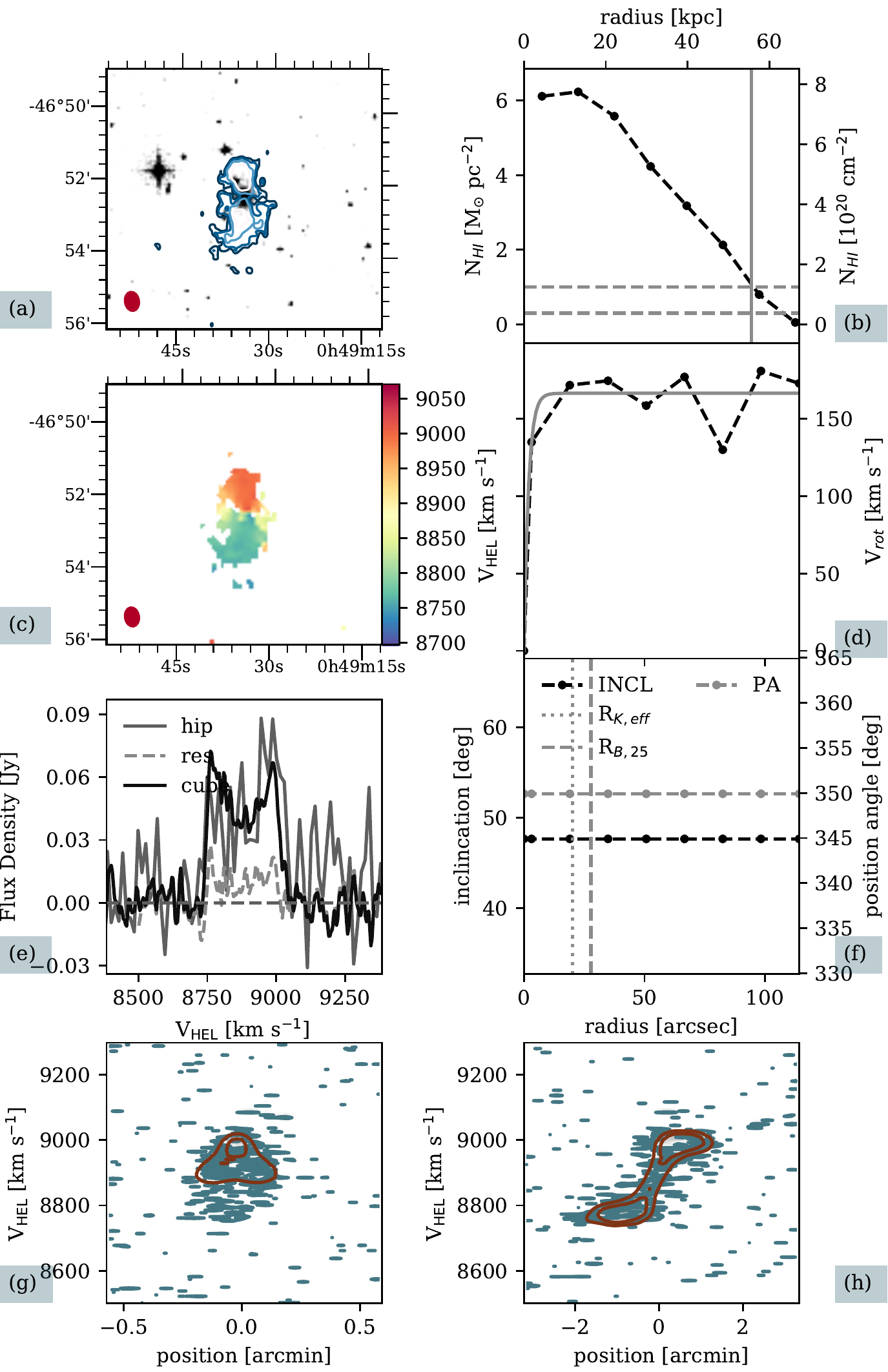}
\caption{ESO243-G002}
\end{figure*}

\begin{figure*}
\includegraphics[width=5.5in]{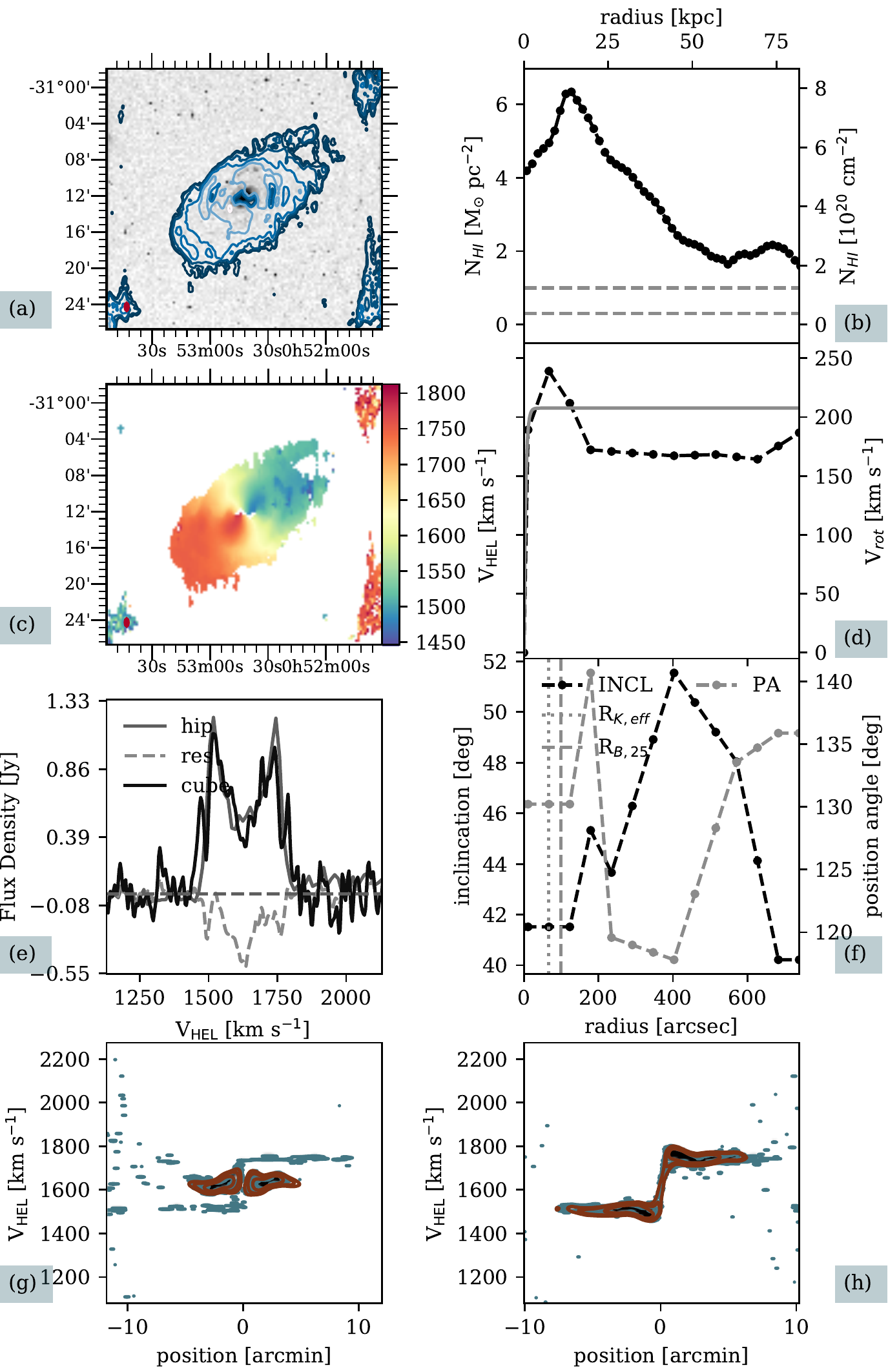}
\caption{NGC289}
\end{figure*}

\begin{figure*}
\includegraphics[width=5.5in]{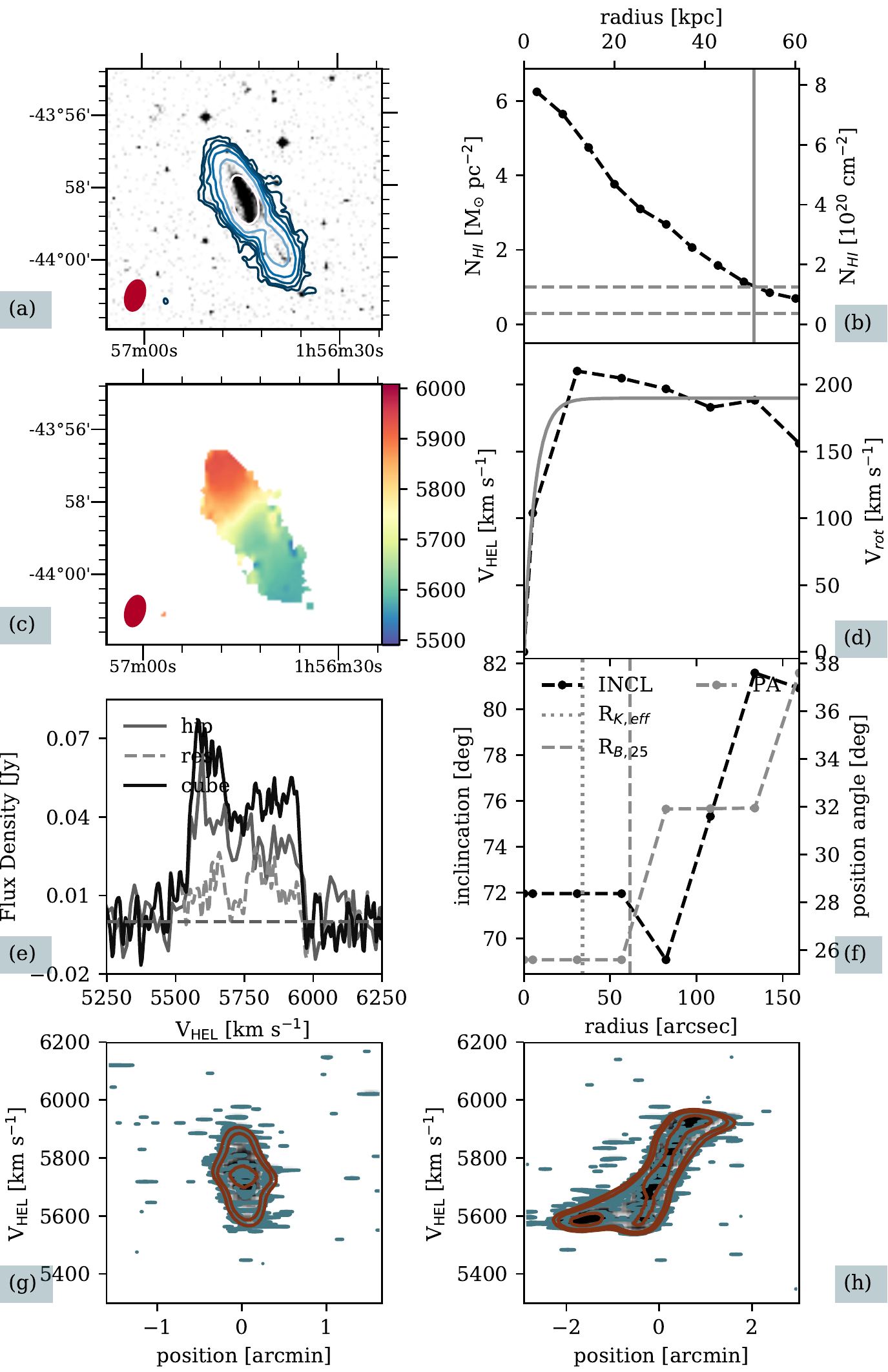}
\caption{ESO245-G010}
\end{figure*}

\begin{figure*}
\includegraphics[width=5.5in]{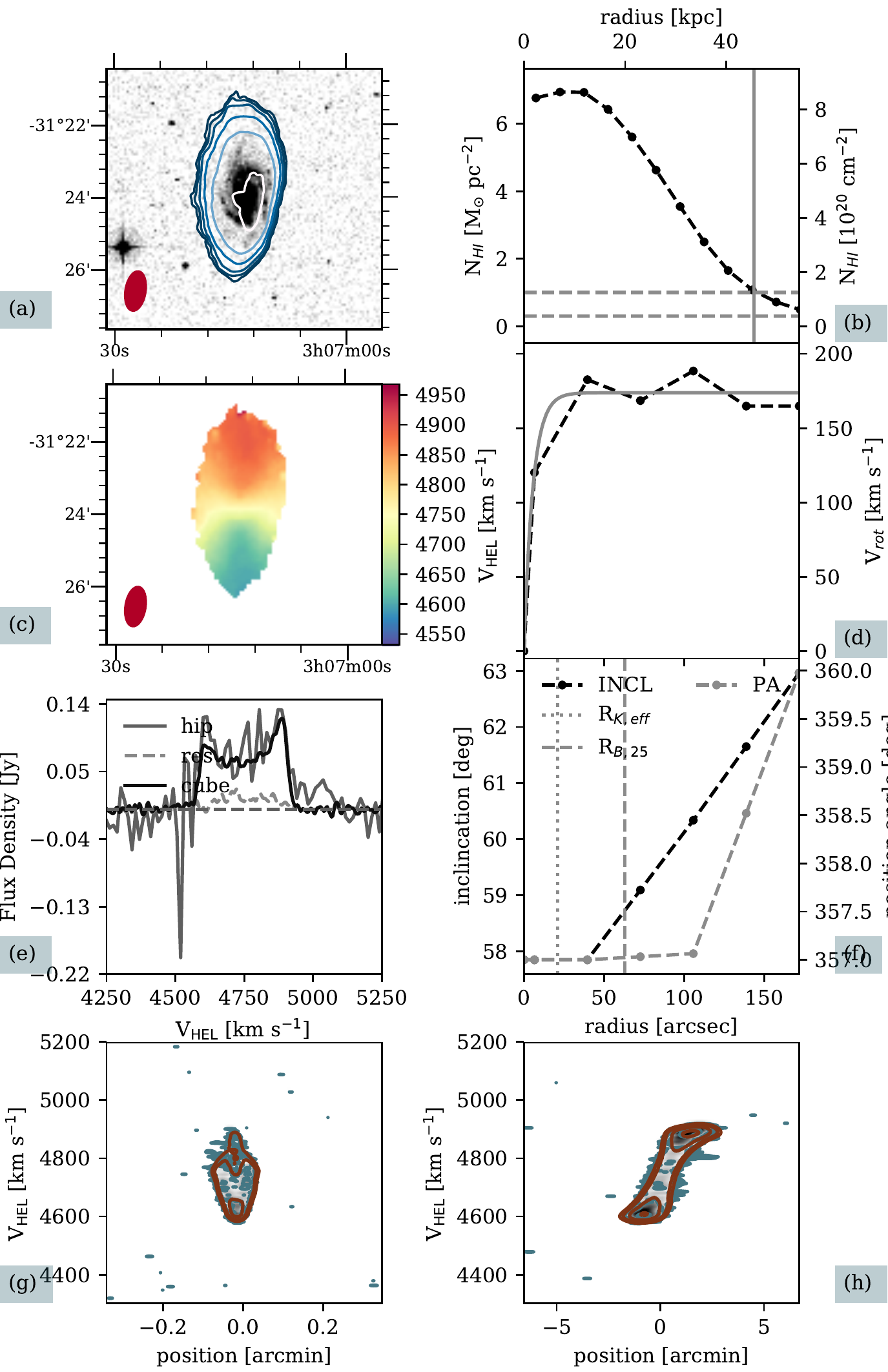}
\caption{ESO417-G018}
\end{figure*}

\begin{figure*}
\includegraphics[width=5.5in]{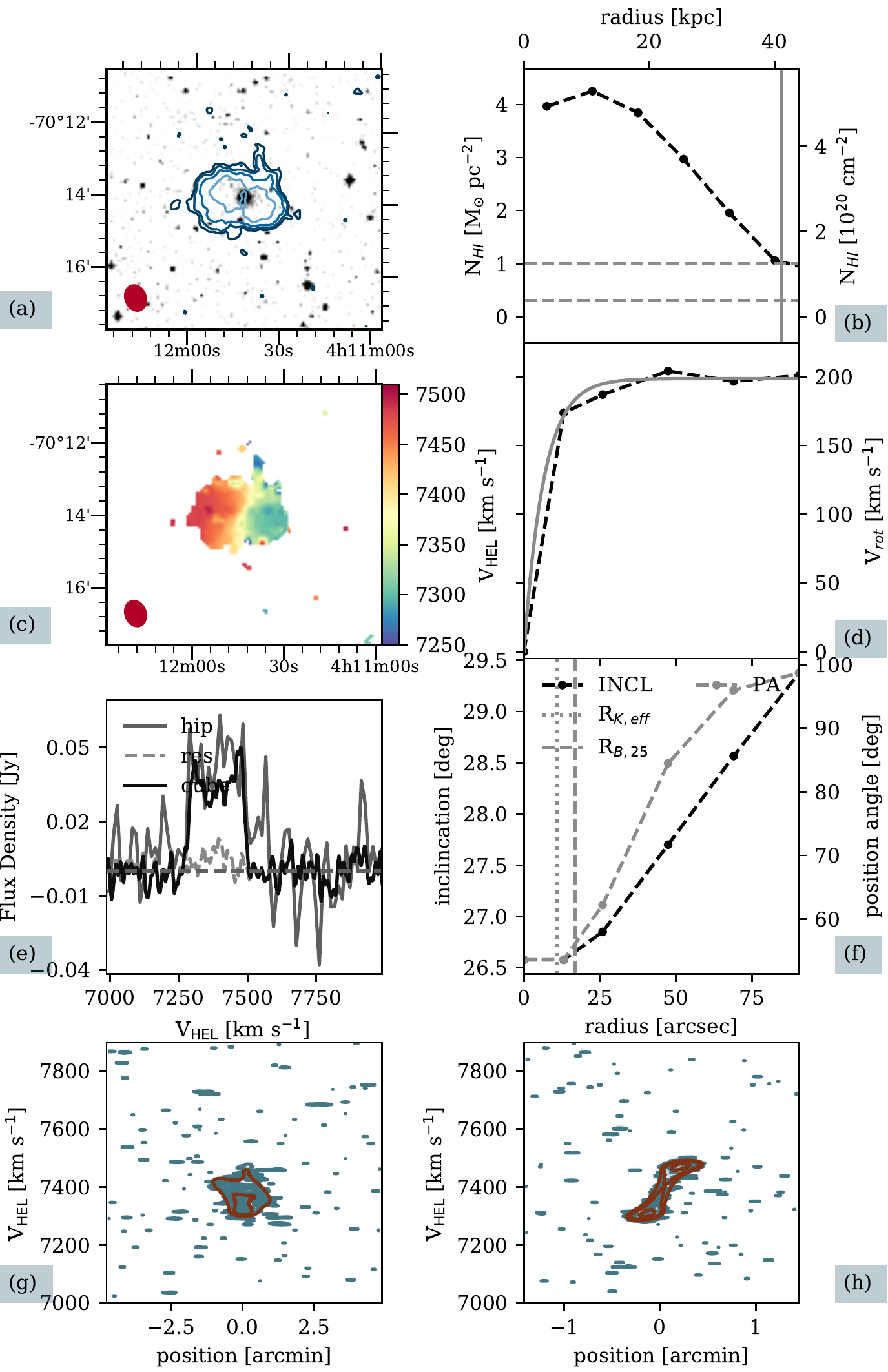}
\caption{ESO055-G013}
\end{figure*}

\begin{figure*}
\includegraphics[width=5.5in]{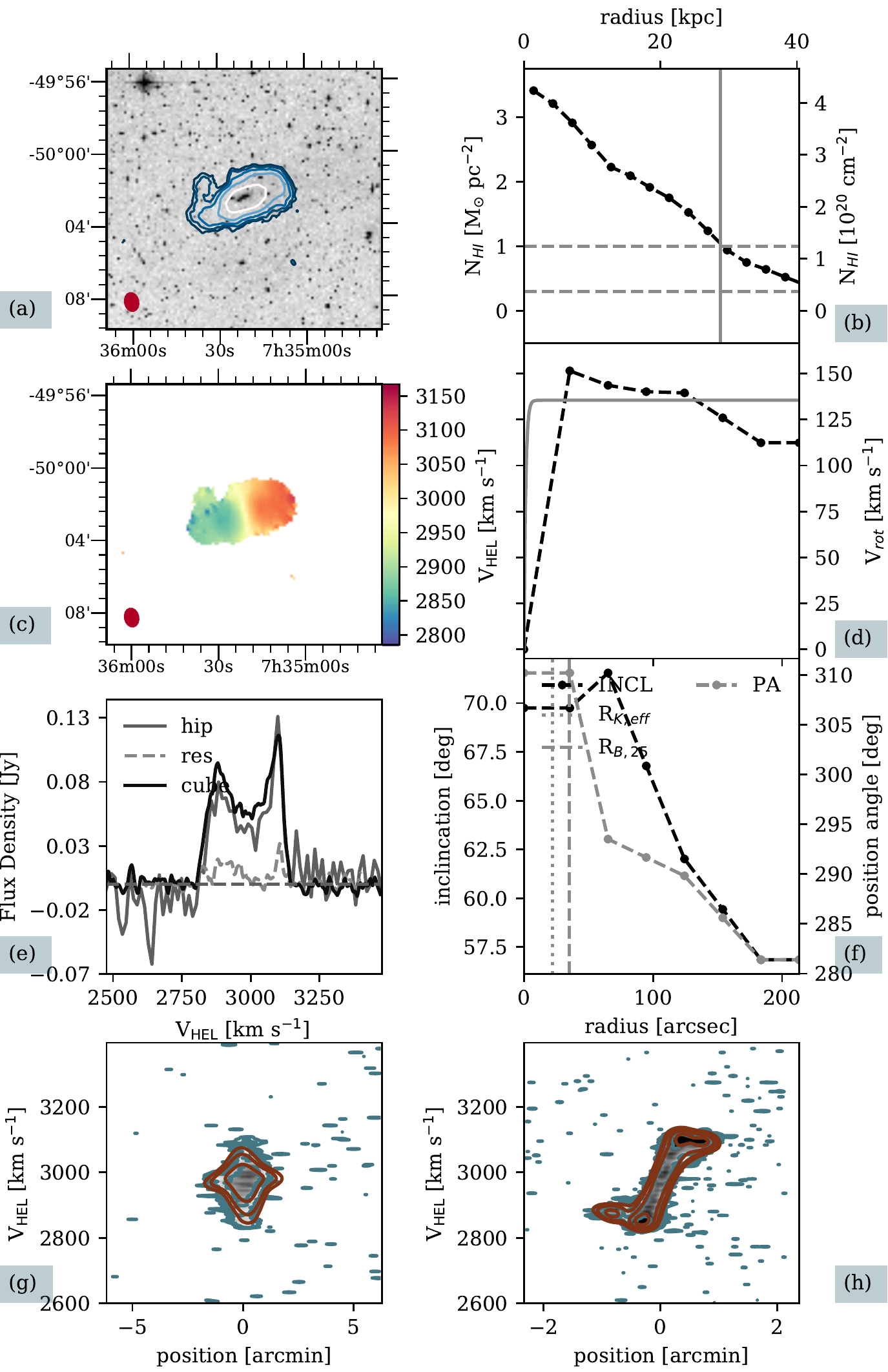}
\caption{ESO208-G026}
\end{figure*}

\begin{figure*}
\includegraphics[width=5.5in]{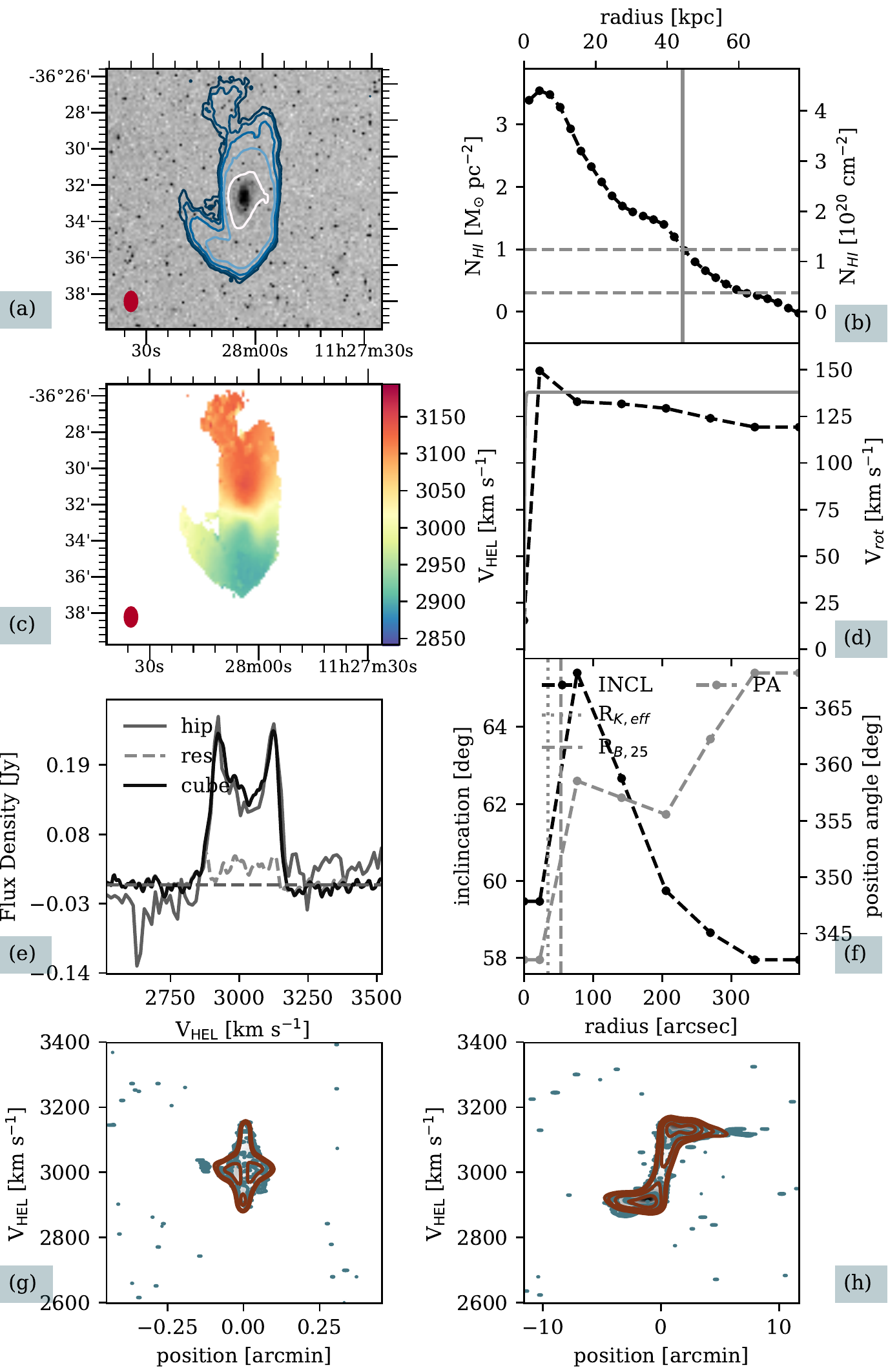}
\caption{ESO378-G003}
\end{figure*}

\begin{figure*}
\includegraphics[width=5.5in]{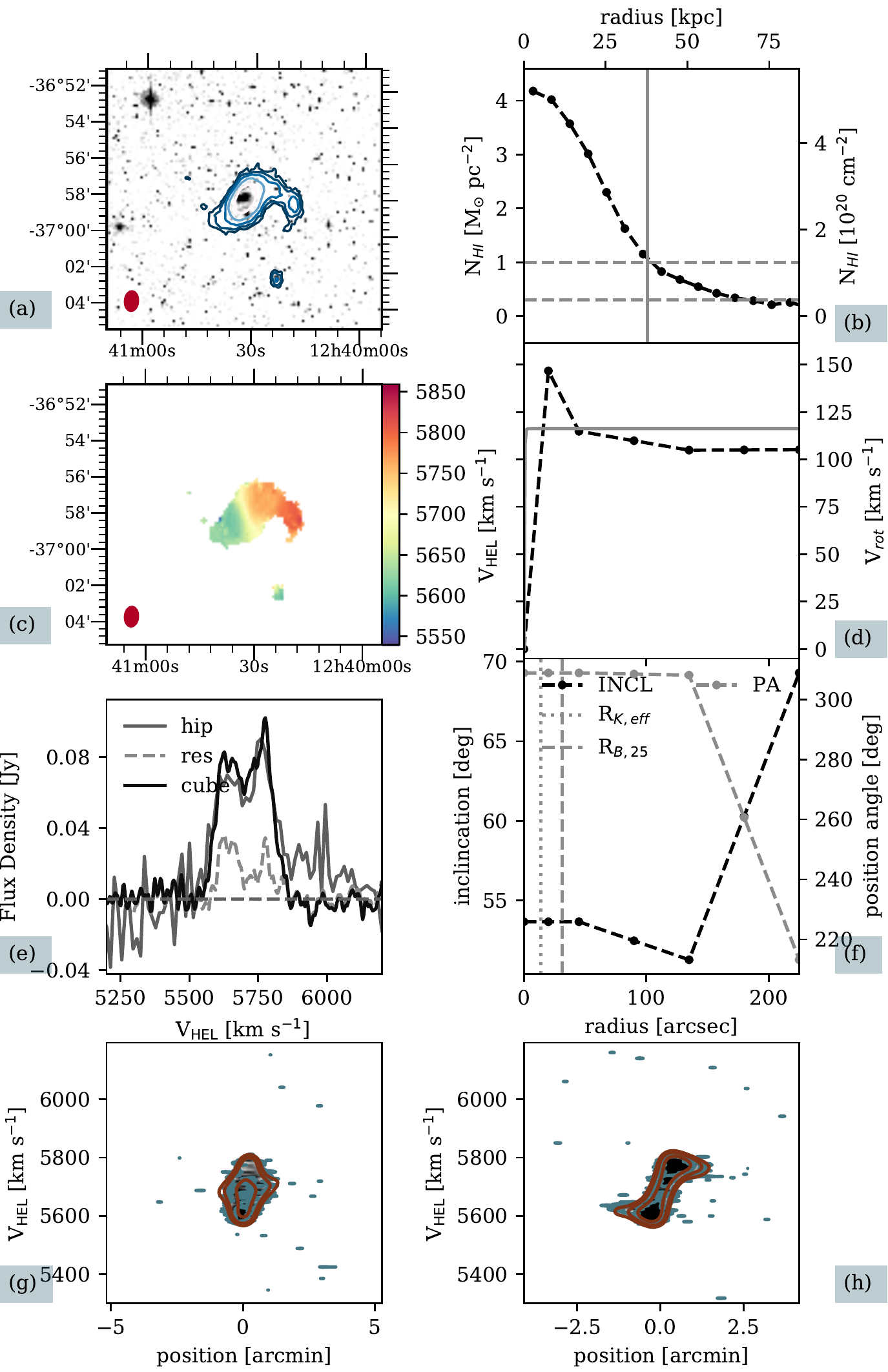}
\caption{ESO381-G005}
\end{figure*}

\begin{figure*}
\includegraphics[width=5.5in]{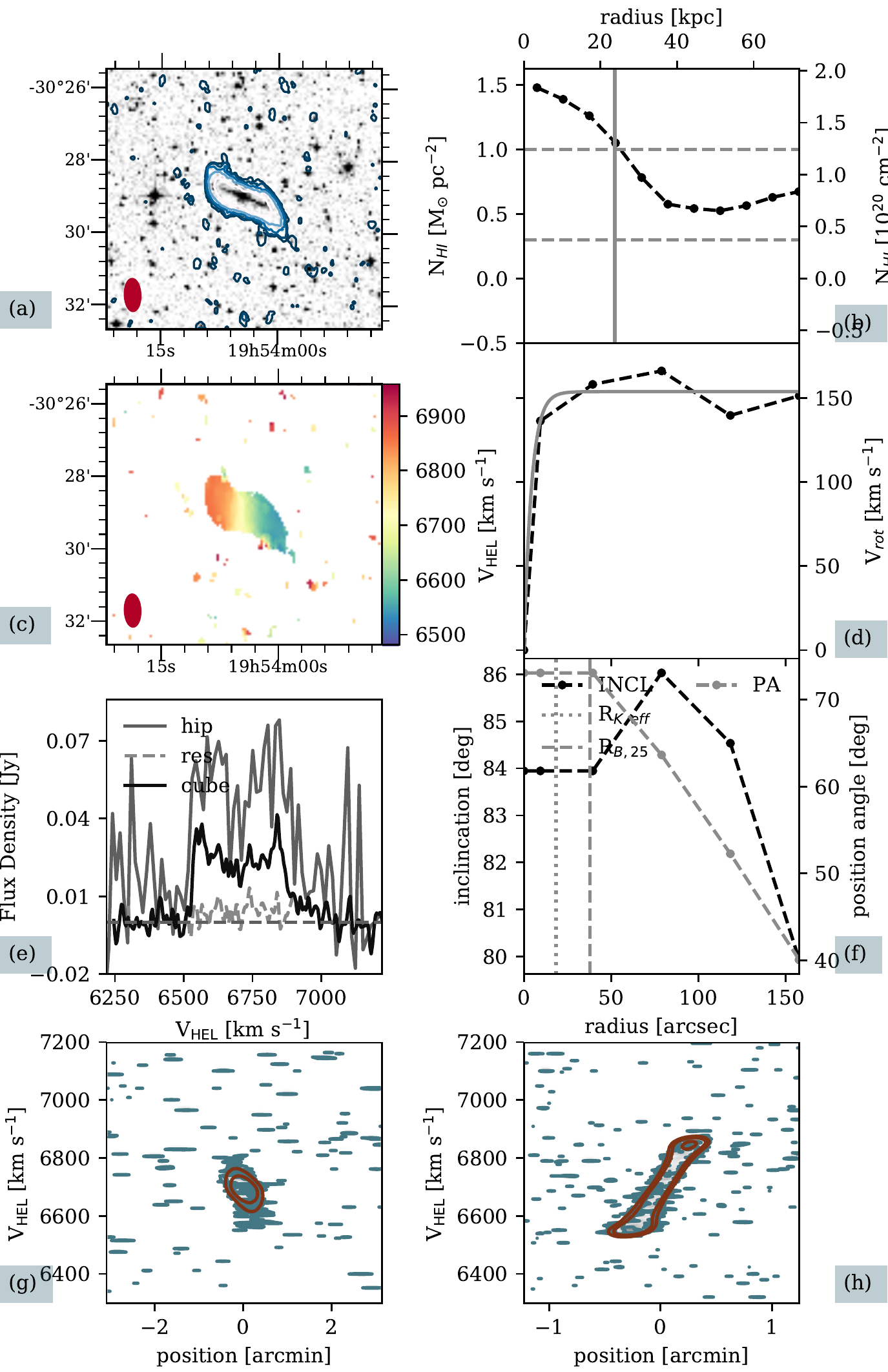}
\caption{ESO461-G010}
\end{figure*}

\begin{figure*}
\includegraphics[width=5.5in]{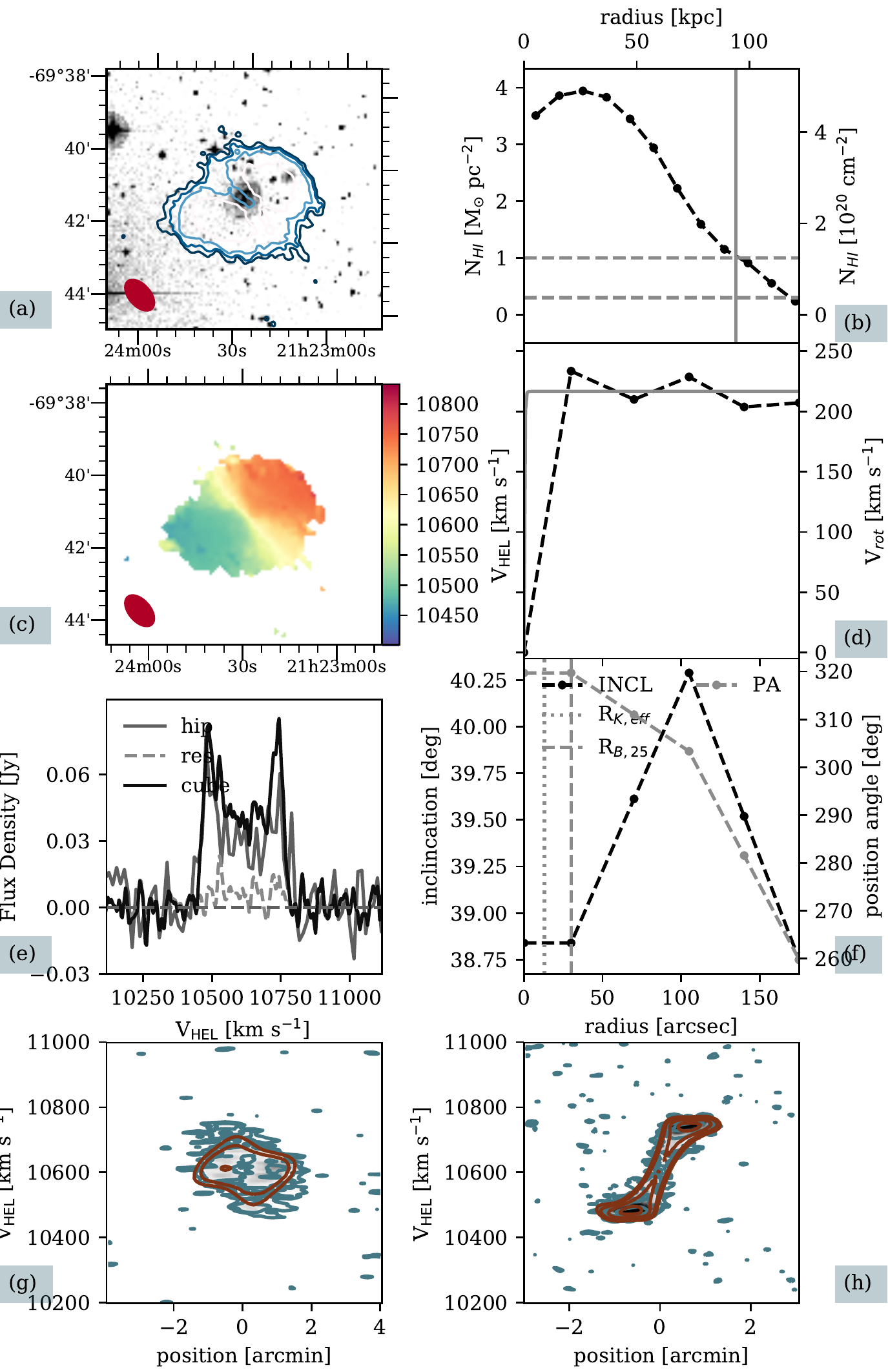}
\caption{ESO075-G006}
\end{figure*}

\begin{figure*}
\includegraphics[width=5.5in]{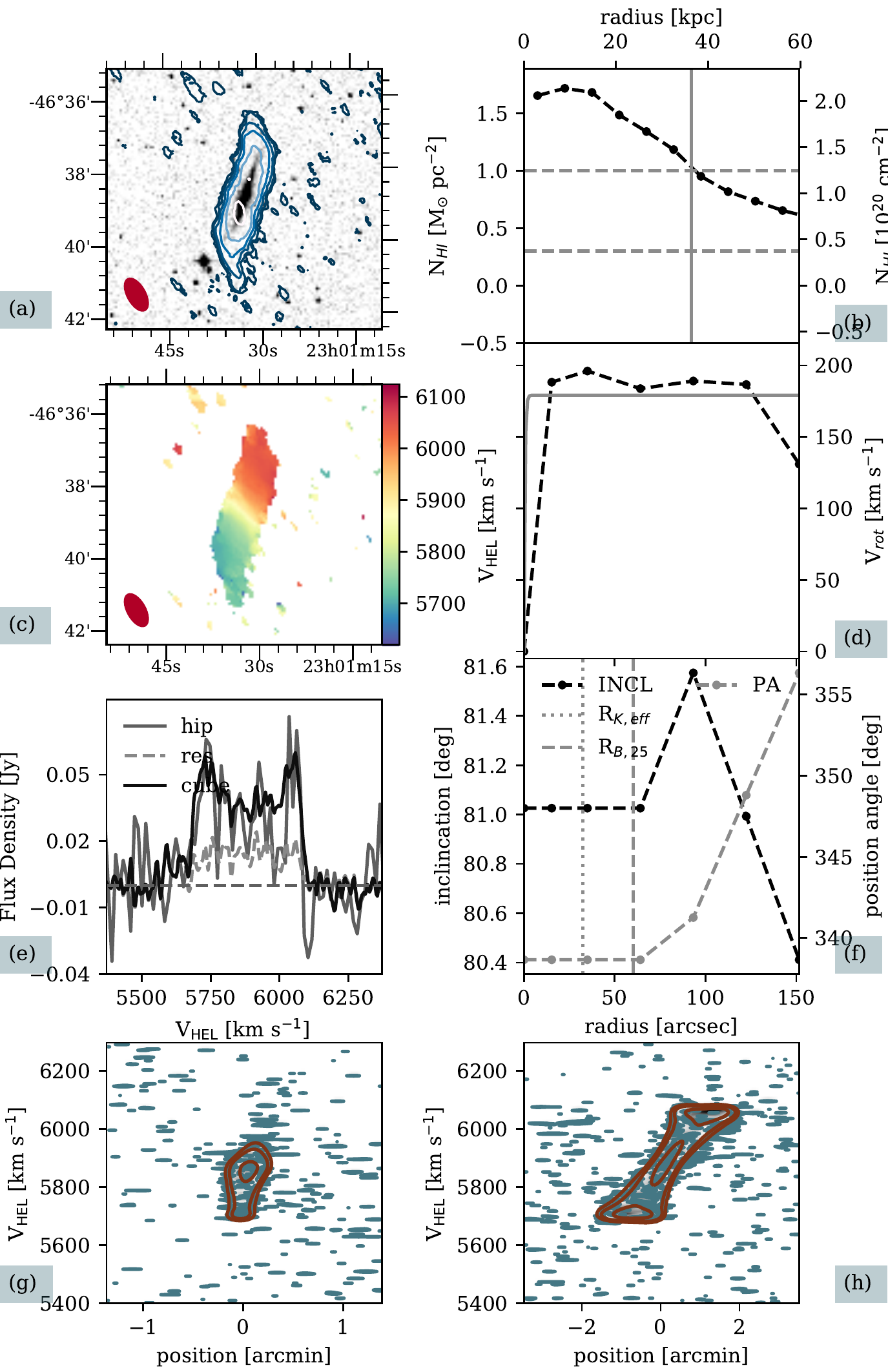}
\caption{ESO290-G035}
\end{figure*}

\newpage

\section{\change{ATCA HI data }of the \control\ sample}
\label{app:con}

This section presents the data panels of the \control\ galaxies. The panels are 
structured as in App.~\ref{app:hix}. 
% \clearpage
\begin{figure*}
\includegraphics[width=5.5in]{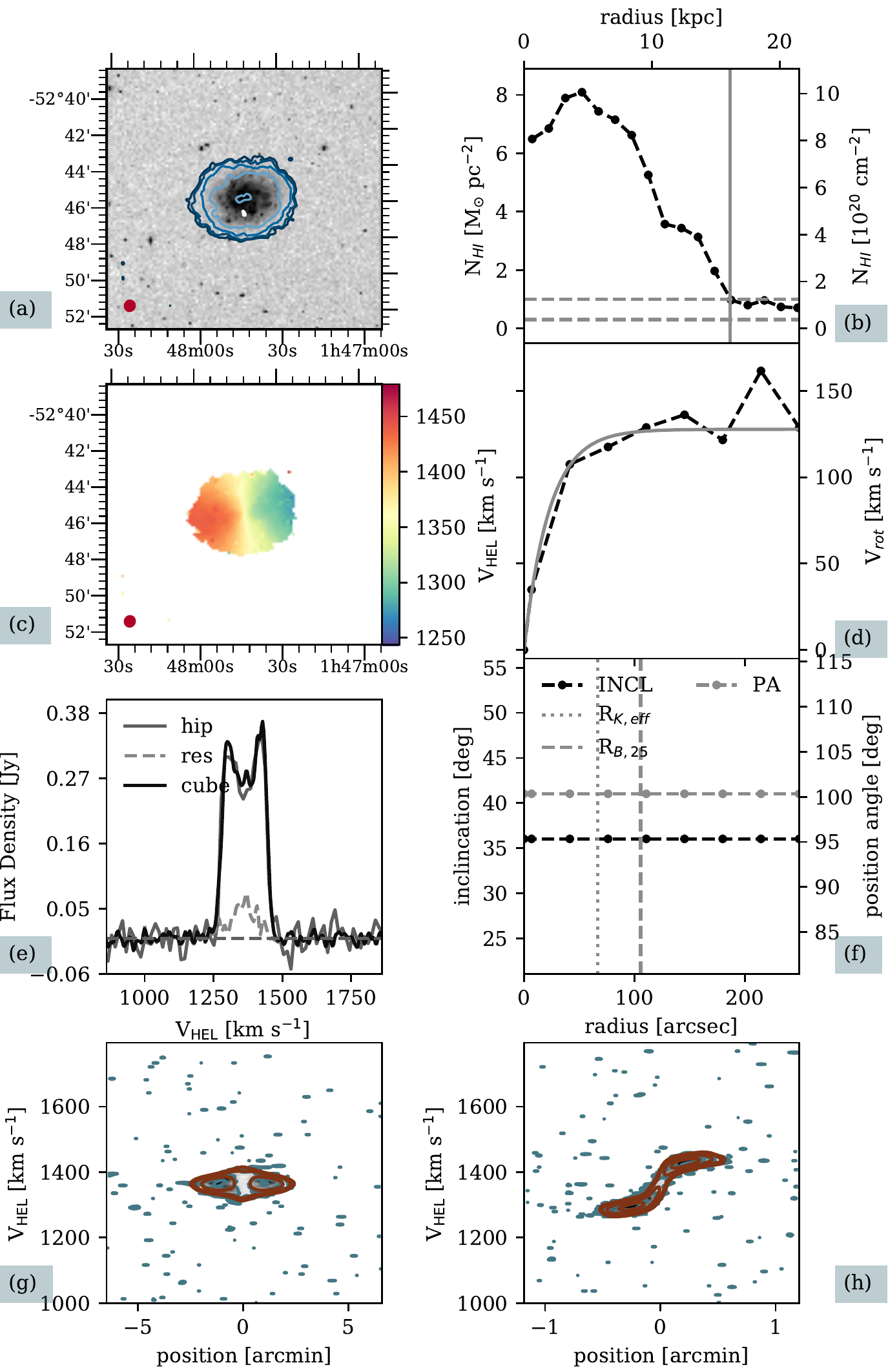}
\caption{NGC685}
\end{figure*}

\begin{figure*}
\includegraphics[width=5.5in]{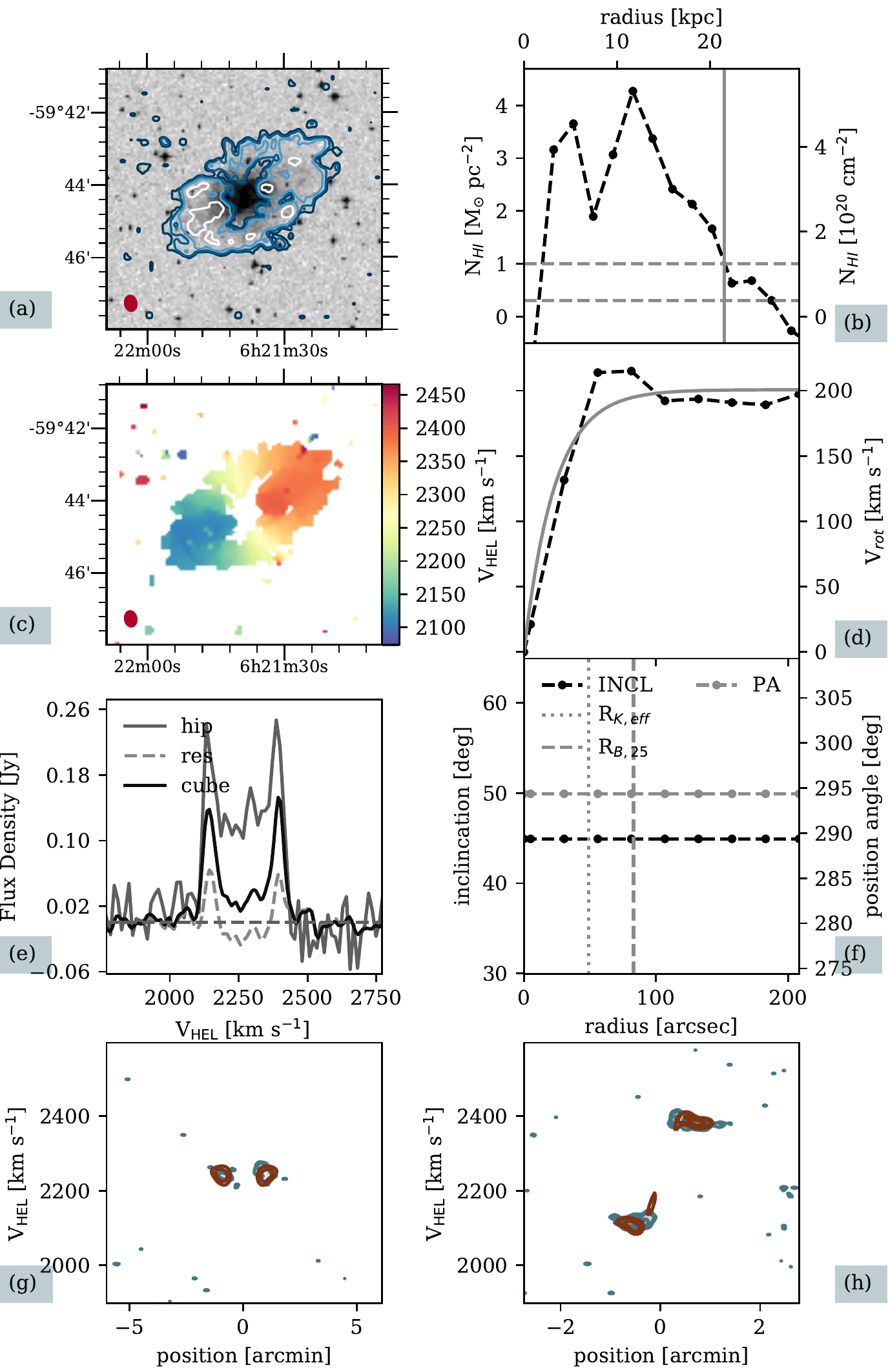}
\caption{ESO121-G026}
\end{figure*}

\begin{figure*}
\includegraphics[width=5.5in]{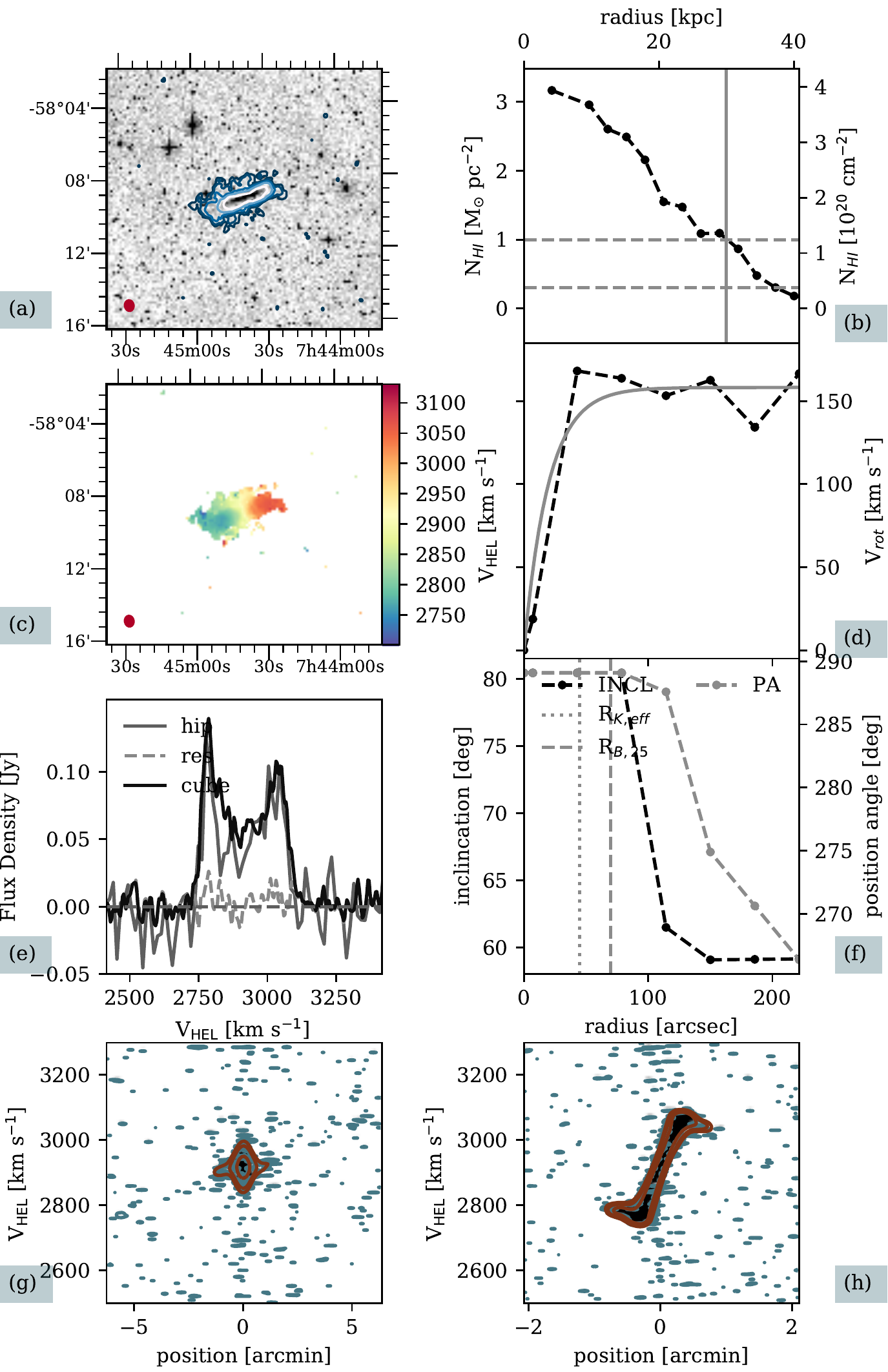}
\caption{ESO123-G023}
\end{figure*}

\begin{figure*}
\includegraphics[width=5.5in]{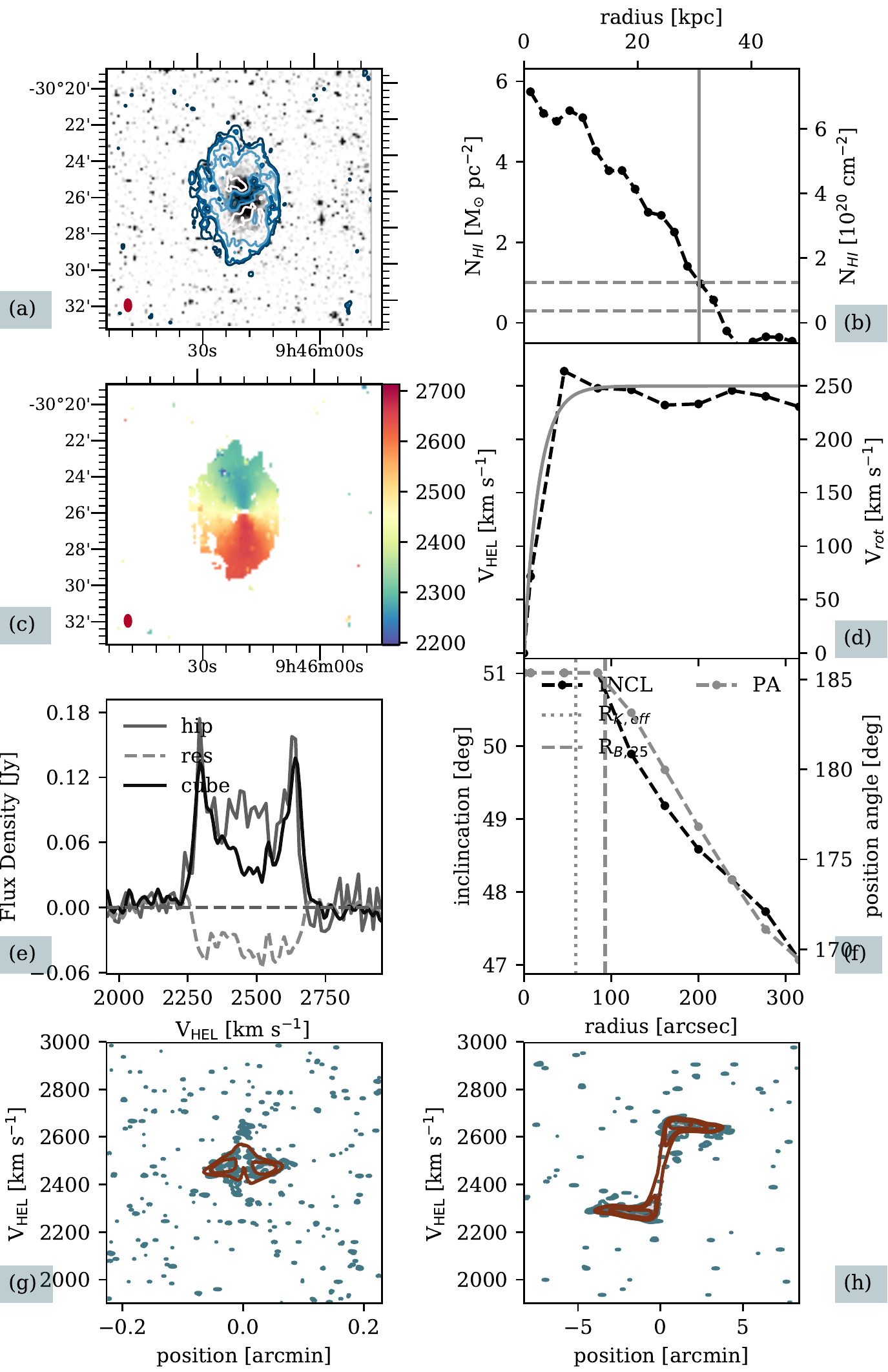}
\caption{NGC3001}
\end{figure*}

\begin{figure*}
\includegraphics[width=5.5in]{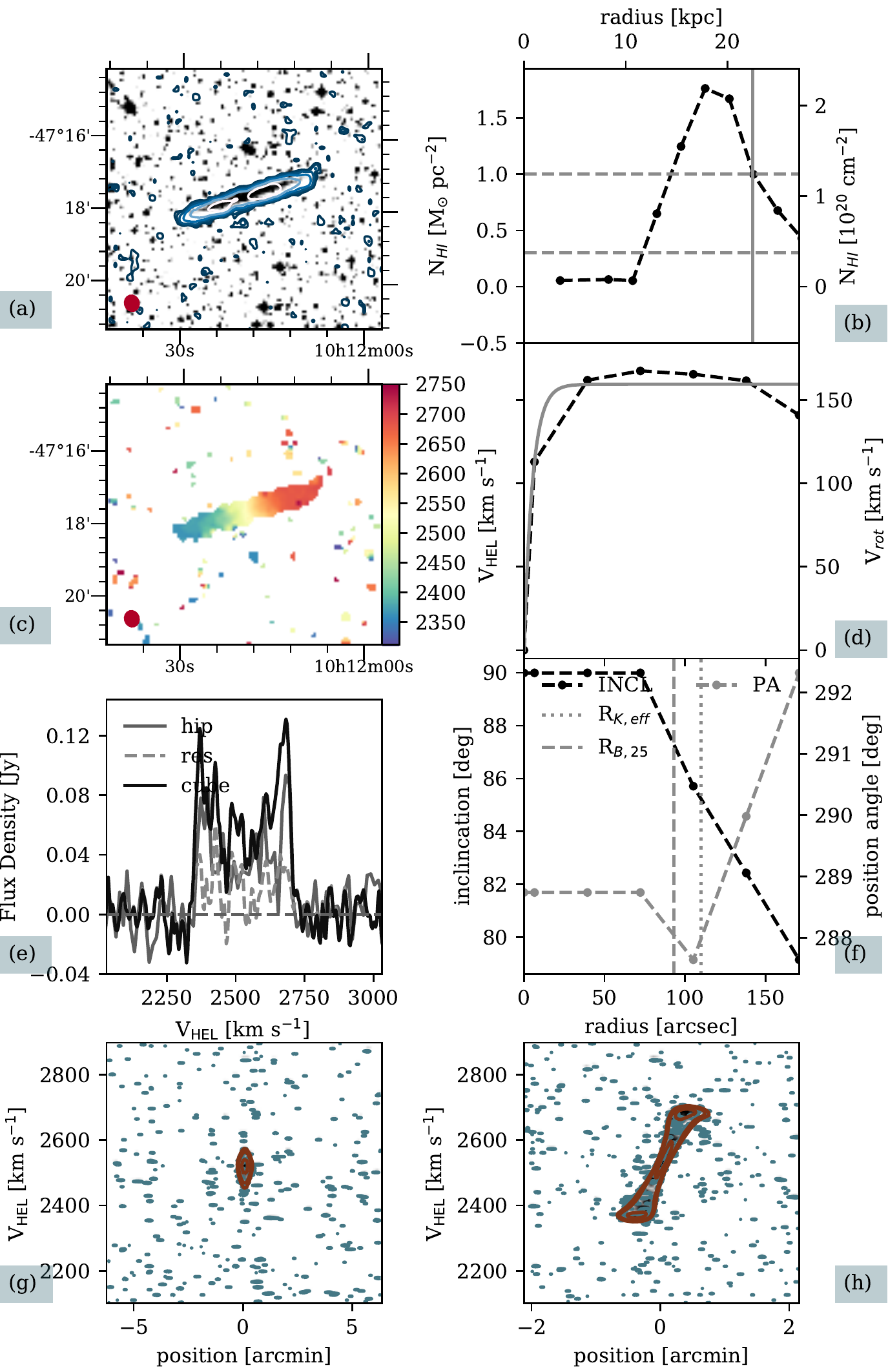}
\caption{ESO263-G015}
\end{figure*}

\begin{figure*}
\includegraphics[width=5.5in]{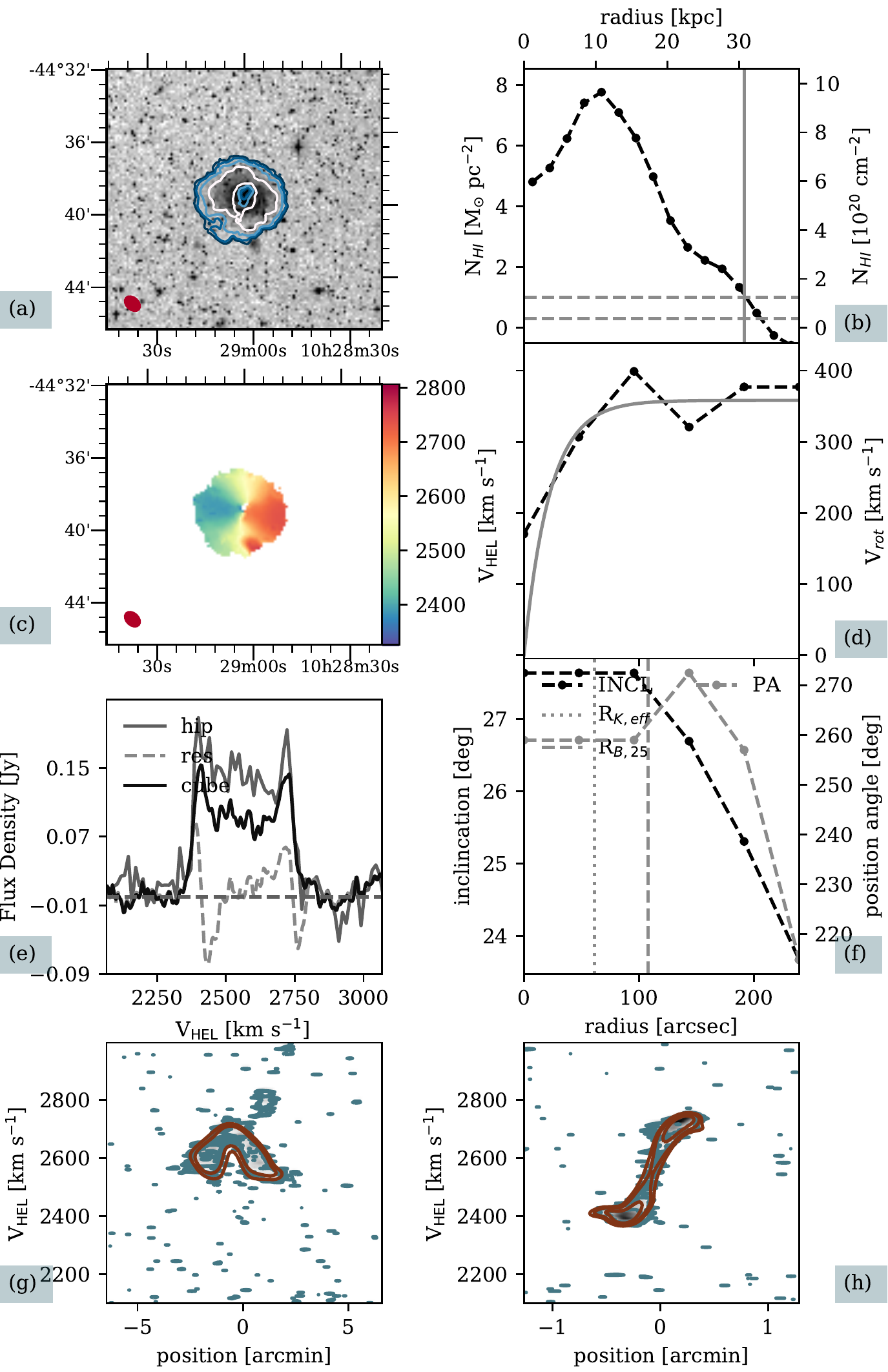}
\caption{NGC3261}
\end{figure*}

\begin{figure*}
\includegraphics[width=5.5in]{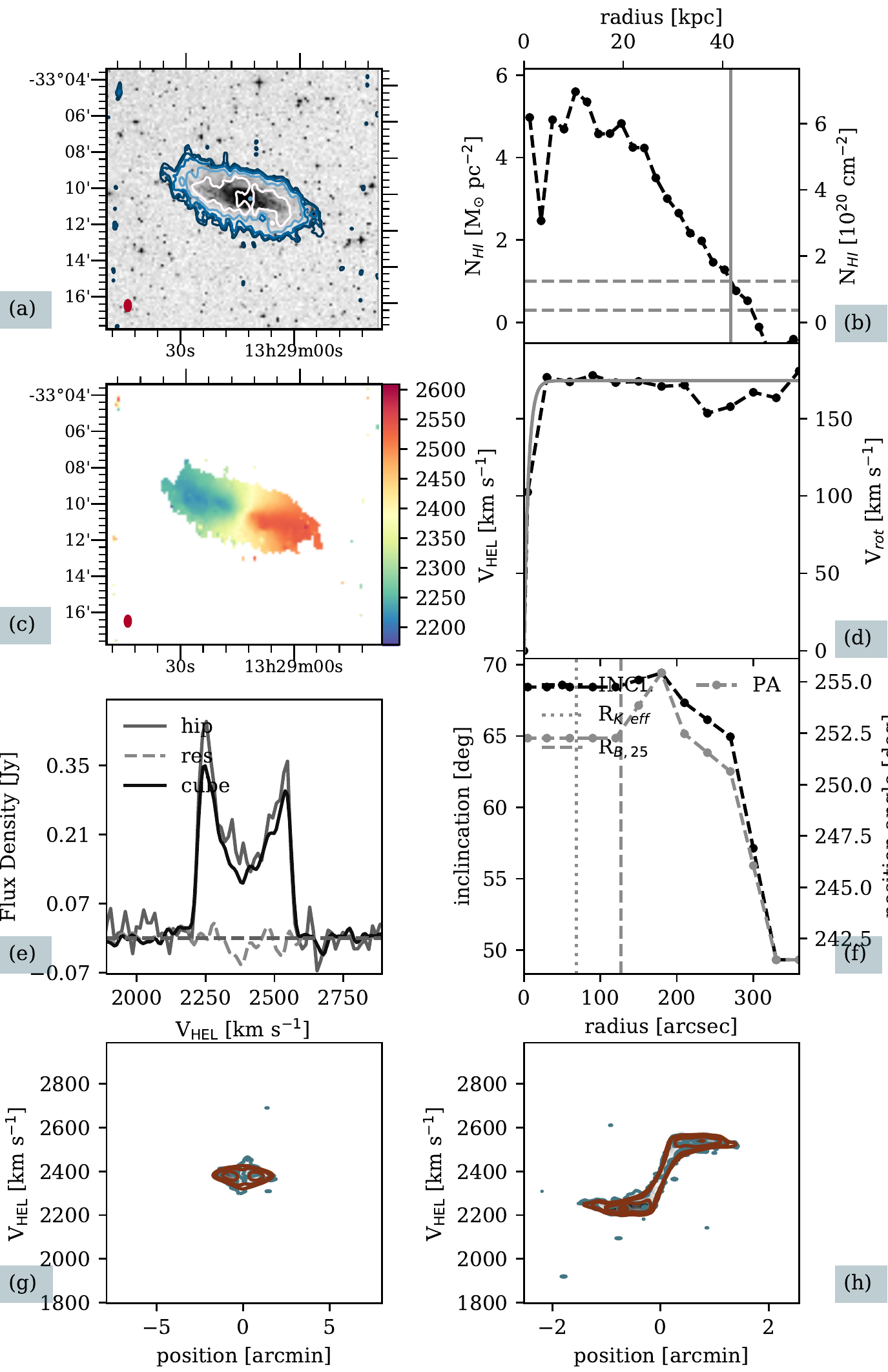}
\caption{NGC5161}
\end{figure*}

\begin{figure*}
\includegraphics[width=5.5in]{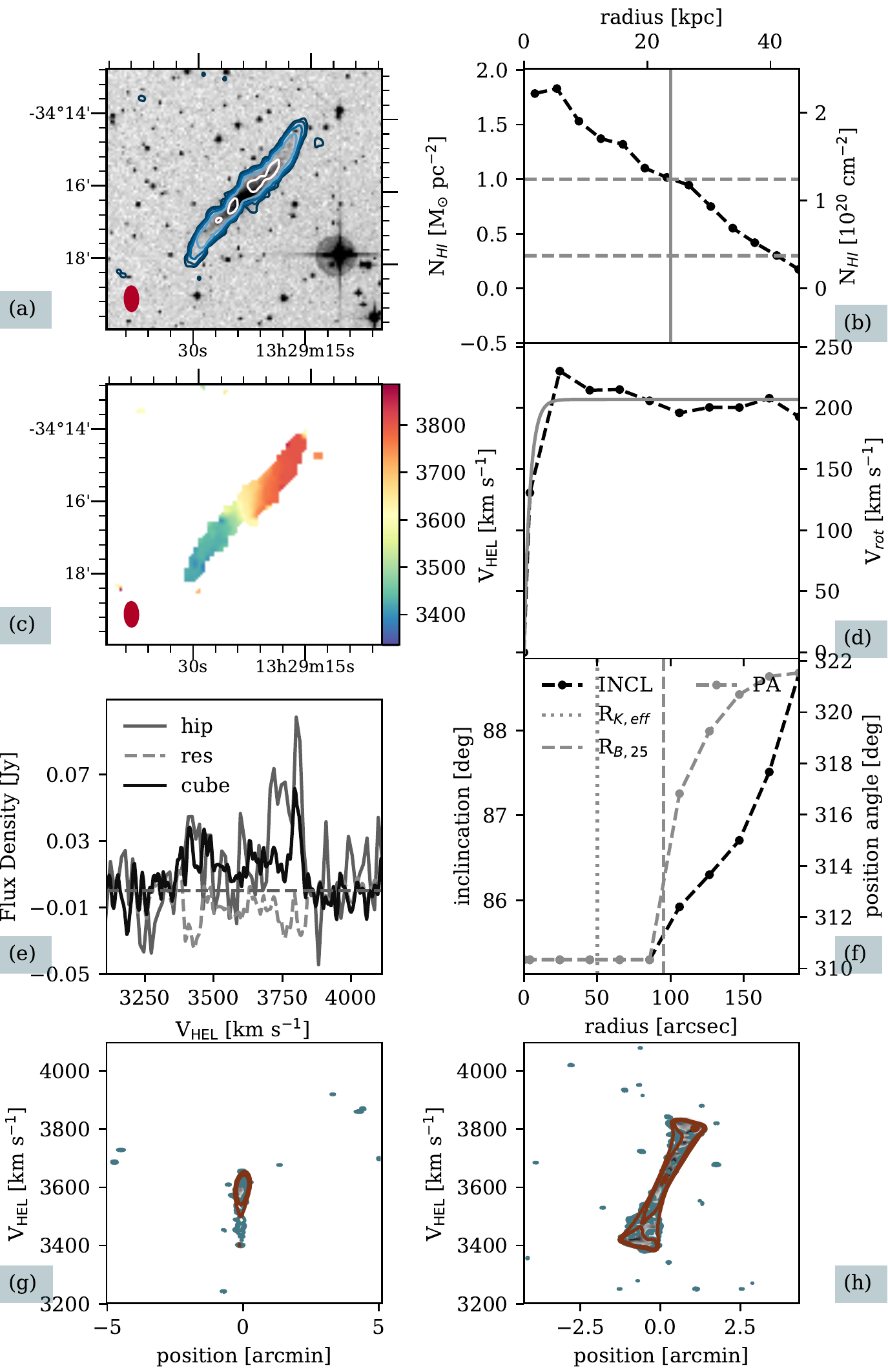}
\caption{ESO383-G005}
\end{figure*}

\begin{figure*}
\includegraphics[width=5.5in]{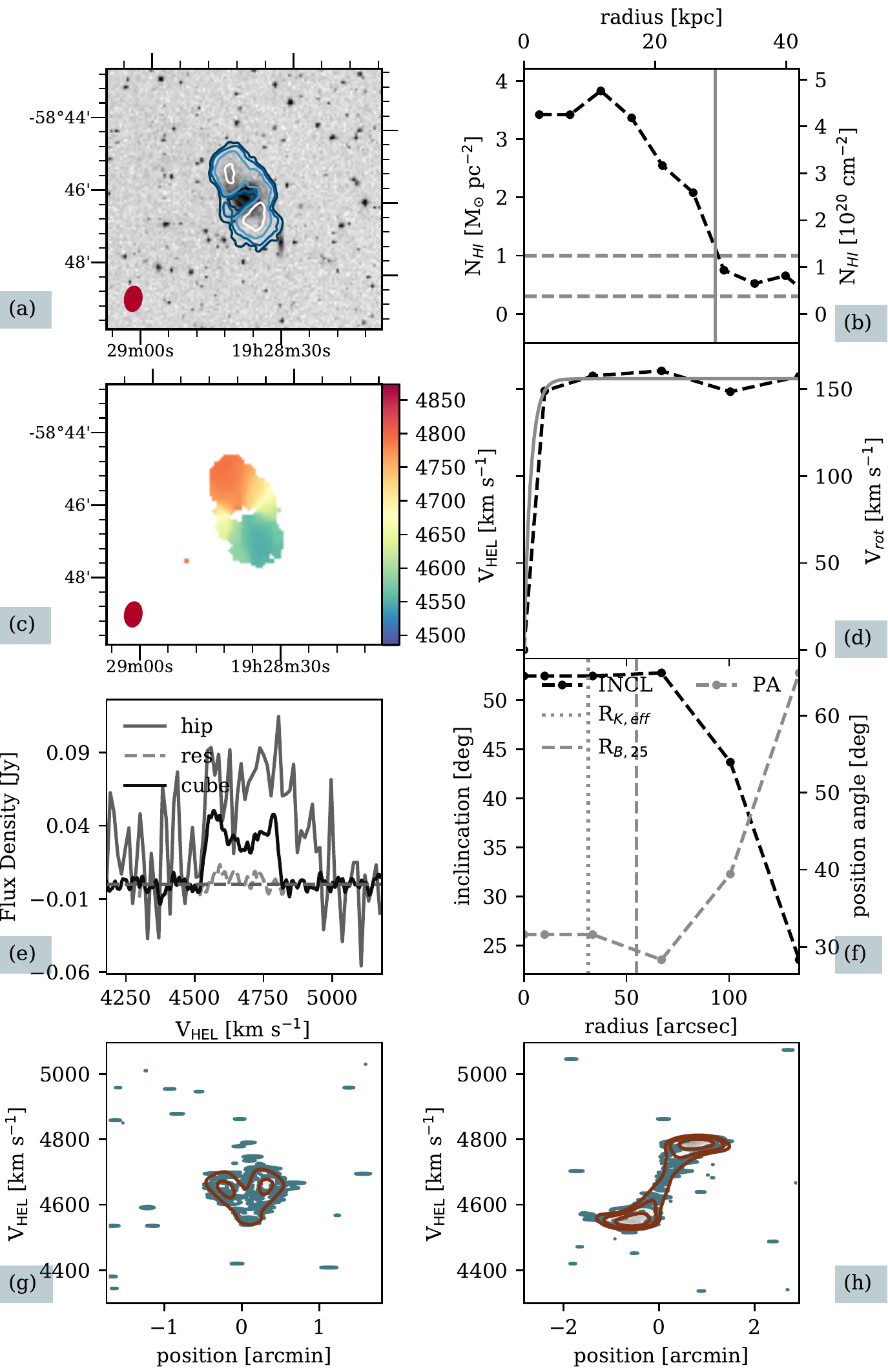}
\caption{IC4857}
\end{figure*}

\begin{figure*}
\includegraphics[width=5.5in]{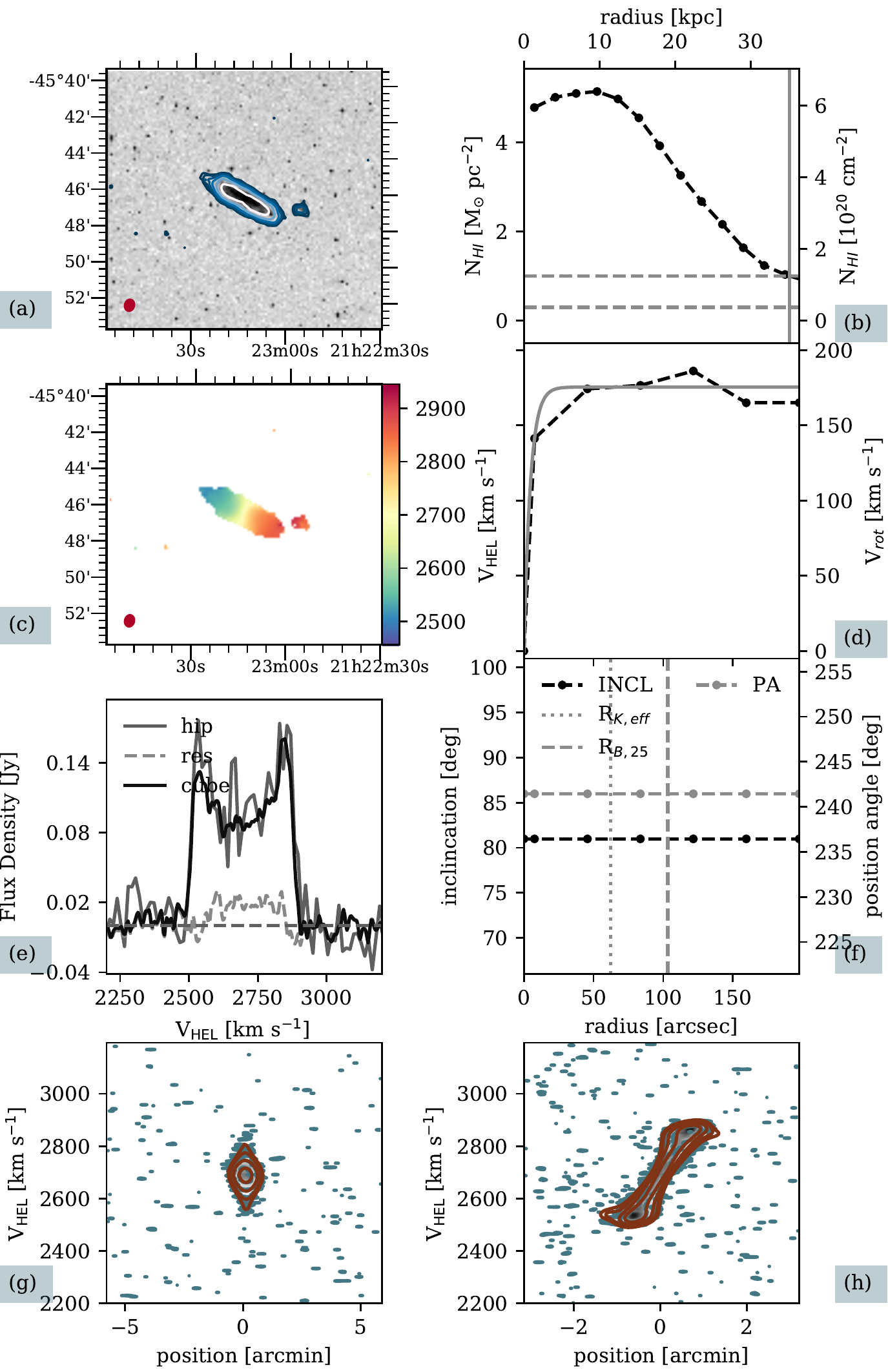}
\caption{ESO287-G013}
\end{figure*}

\begin{figure*}
\includegraphics[width=5.5in]{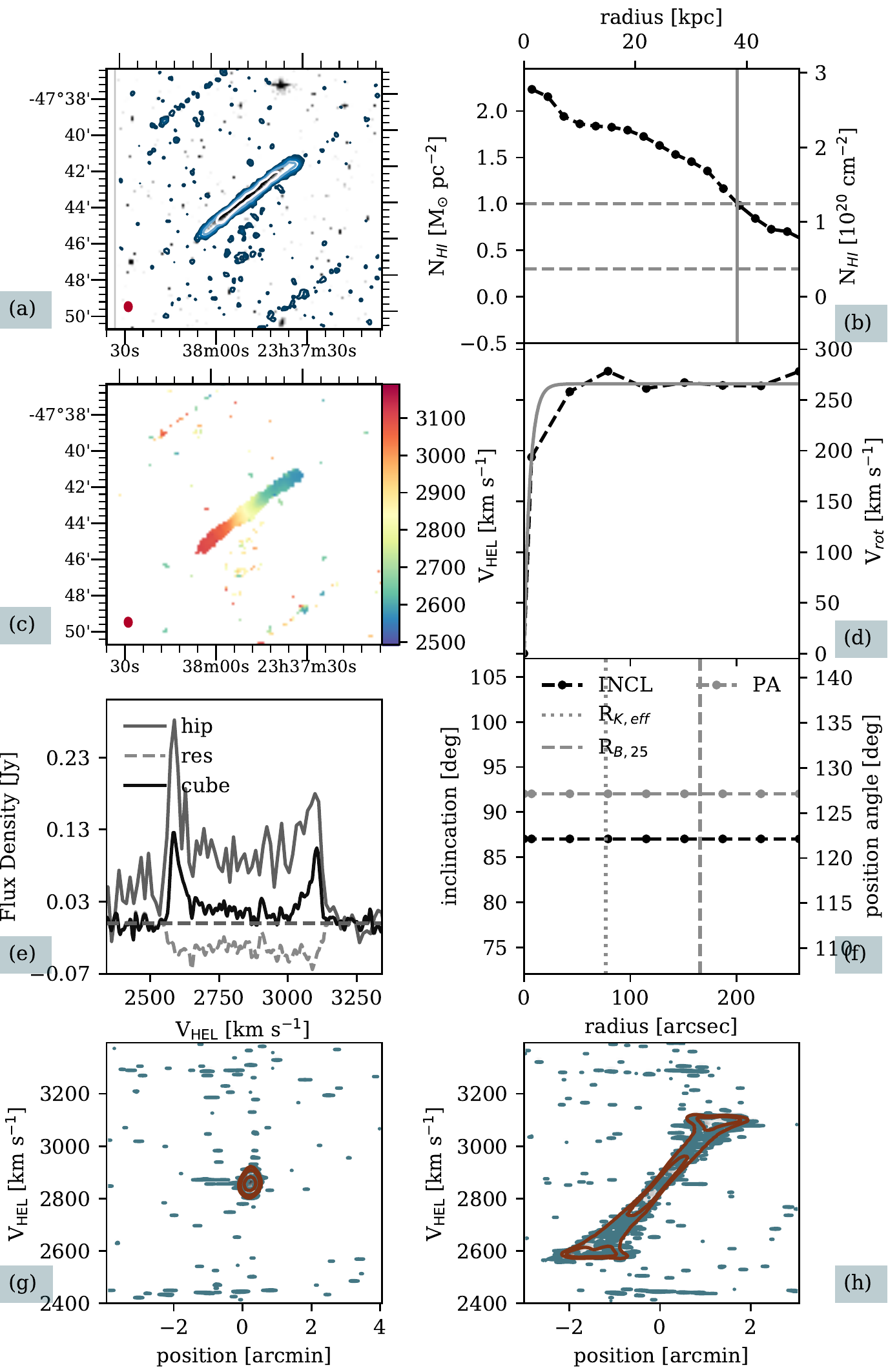}
\caption{ESO240-G011}
\end{figure*}

\bsp

\label{lastpage}

\end{document}